\documentclass[11pt]{article}
\pdfoutput=1

\usepackage{amsmath}
\usepackage{amssymb}
\usepackage{mathtools}
\usepackage{amsthm}
\usepackage{tabularx}
\usepackage{float}

\usepackage{microtype}
\usepackage{graphicx}
\usepackage{subfigure}
\usepackage{booktabs} %
\usepackage{amssymb}
\usepackage{bm}
\usepackage{bbm}
\usepackage{amsmath}  %
\usepackage{amsthm}
\usepackage{xcolor}
\usepackage{textcomp}
\usepackage{multirow}
\usepackage{mathtools}
\usepackage[margin=1in]{geometry}
\usepackage{authblk}
\usepackage[numbers, sort]{natbib}
\usepackage{xspace}
\usepackage{comment}
\usepackage{thmtools, thm-restate}  %
\usepackage{verbatim}

\usepackage{tikz}
\usetikzlibrary{arrows}
\usetikzlibrary{positioning}

\tikzset{
  treenode/.style = {align=center, inner sep=0pt, text centered,
    font=\sffamily},
  arn_n/.style = {treenode, circle, black, font=\sffamily\bfseries, draw=black,
    fill=white, text width=1.5em},%
  arn_r/.style = {treenode, circle, black, font=\sffamily\bfseries, draw=black,
    fill=white, text width=1.0em},%
  arn_x/.style = {treenode, rectangle, draw=black,
    minimum width=0.5em, minimum height=0.5em}%
}

\usepackage{hyperref}

\usepackage{bm}
\usepackage{enumitem}

\newcommand{\parastart}[1]{\textbf{#1~~~}}

\newcommand*{\ShowNotes}{}

\newcommand{\seqlen}{\mathbf{S}}

\ifdefined\ShowNotes
  \newcommand{\colornote}[3]{{\color{#1}\bf{#2 #3}\normalfont}}
\else
  \newcommand{\colornote}[3]{}
\fi

\definecolor{darkred}{rgb}{0.7,0.1,0.1}
\definecolor{darkgreen}{rgb}{0.1,0.5,0.1}
\definecolor{cyan}{rgb}{0.7,0.0,0.7}
\definecolor{dblue}{rgb}{0.2,0.2,0.8}
\definecolor{maroon}{rgb}{0.76,.13,.28}
\definecolor{burntorange}{rgb}{0.81,.33,0}
\definecolor{royalpurple}{rgb}{0.47,.31,0.66}

\ifdefined\ShowNotes
  
\else
  
\fi

\newif\ifarxiv

\title{Benchmarking and Building Long-Context Retrieval Models with LoCo and
M2-BERT}

\author{Jon Saad-Falcon$^\dag$, Daniel Y. Fu$^\dag$, Simran Arora$^\dag$, Neel Guha$^\dag$, Christopher Ré$^\dag$}%

\date{%
    $^\dag$Stanford University, Department of Computer Science\\
    \vspace{0.1cm}
    \texttt{\{jonsaadfalcon, danfu, simarora, nguha, chrismre\}@stanford.edu}\\
    \vspace{0.5cm}
    \today
}

\begin{document}

\maketitle

\begin{abstract}

Retrieval pipelines---an integral component of many machine learning systems---perform poorly in domains where documents are long (e.g., 10K tokens or more) and where identifying the relevant document requires synthesizing information across the entire text. Developing long-context retrieval encoders suitable for these domains raises three challenges: (1) how to evaluate long-context retrieval performance, (2) how to pretrain a base language model to represent both short contexts (corresponding to queries) and long contexts (corresponding to documents), and (3) how to finetune this model for retrieval under the batch size limitations imposed by GPU memory constraints. To address these challenges, we first introduce LoCoV1, a novel 12 task benchmark constructed to measure long-context retrieval where chunking is not possible or not effective. We next present the M2-BERT retrieval encoder, an 80M parameter state-space encoder model built from the Monarch Mixer architecture, capable of scaling to documents up to 32K tokens long. We describe a pretraining data mixture which allows this encoder to process both short and long context sequences, and a finetuning approach that adapts this base model to retrieval with only single-sample batches. Finally, we validate the M2-BERT retrieval encoder on LoCoV1, finding that it outperforms competitive Transformer-based models by at least 23.3 points, despite containing upwards of 90$\times$ fewer parameters. 
\end{abstract}

\section{Introduction}
\label{introduction}

Retrieval is an essential component of machine learning pipelines for tasks like search, question-answering, dialogue, and fact verification \cite{chen2017retrieveread, lewis2021retrievalaugmented, dinan2019wow, petroni-etal-2021-kilt}. Most retrieval systems rely on pretrained text models that are only capable of processing short input sequences (e.g., approximately 512 to 8192 tokens) \cite{Reimers2019SentenceBERTSE, 10.1145/3477495.3531833, karpukhin-etal-2020-dense, santhanam-etal-2022-colbertv2}. Yet from our analysis of domain-specific datasets, such as those in law and medicine (Section~\ref{sec:M2_quality}), the documents or queries may be tens of thousands of tokens long, and identifying the relevant document requires synthesizing information across a long text sequence~\cite{li2023don}.
Examples include legal contracts, company financial documents, patient notes, screenplays, and other documents with specific contextual details and cross-document references ~\cite{bai2023longbench, shaham2022scrolls, dasigi2021dataset, xu2023retrieval}. Our work explores how to benchmark and build high quality and efficient retrieval systems for long-context corpora. 

Popular retrieval models are built using the Transformer architecture \cite{vaswani2023attention}, which scales quadratically in sequence length, making it expensive to extend existing retrieval recipes to the long-context setting. 
Recent work on \textit{state-space} architectures, such as S4~\cite{gu2022efficiently}, Mamba~\cite{gu2023mamba}, Monarch Mixer~\cite{fu2023monarch}, and more~\cite{wang2022pretraining, smith2023simplified, hasani2022liquid, fu2023hungry, poli2023hyena}, suggests that the subquadratic scaling properties enjoyed by these models make them amenable for long contexts.
However, adapting state-space models for retrieval raises three challenges:

\begin{figure*}[t]
   \centering
   \includegraphics[width=\linewidth]{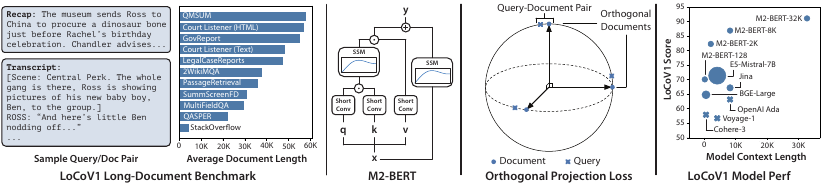}
   \caption{\textbf{Left:} The LoCoV1 long document retrieval benchmark and the average document length of its constituent datasets. \textbf{Center Left:} M2-BERT sequence mixer. \textbf{Center Right:} The orthogonal projection loss. \textbf{Right:} Performance of various retrieval models and M2-BERT at different sequence lengths on LoCoV1. Circles are open models, where circle area corresponds to model size. X marks are closed models.}
   \label{fig:main_figure}
\end{figure*}

\begin{itemize}[leftmargin=*,nosep,nolistsep]
    \item \textbf{Evaluation}: Existing benchmarks for retrieval contain query-document pairs where the relevant information is contained either within the first 512 tokens of the document, or within a small sequence of text \cite{thakur2021beir, muennighoff2022mteb}. As a result, naive truncation-based and chunking baselines perform nearly optimally, regardless of the document length. 
    Validating long-context retrievers thus requires benchmarks on which identifying the relevant document requires reasoning across longer spans of text (e.g. medical, financial, or legal documents with many repeated textual phrases amongst in-class documents but key contextual details throughout the document). 
    \item \textbf{Pretraining}: Retrieval encoders must be pretrained to process both short sequences (corresponding to queries) and long sequences (corresponding to documents). Prior work on state-space model pretraining, in contrast, has focused exclusively on tasks requiring pretraining on uniformly shorter textual inputs \cite{fu2023monarch, wang2022pretraining}. 
    In Section~\ref{sec:ablation}, we show that naive pretraining strategies for model weight initialization are insufficient for preparing retrieval encoders to long input sequences.
    \item \textbf{Finetuning}: Retrieval encoders are usually finetuned from pretrained models using a contrastive loss function (multiple negatives ranking loss, known as MNRL).
    MNRL treats other in-batch positive passages as negative passages for a given query \cite{Reimers2019SentenceBERTSE}.
    MNRL then pushes the embeddings of the positive pair and passage together, while pushing apart the embeddings of negative passages.
    With a  large batch size, this loss creates an embedding geometry that aligns positive pairs together, while distributing them uniformly around the embedding hypersphere~\cite{wang2022understanding, leszczynski2022tabi, henderson2017efficient}
    Training long-context models with the requisite batch sizes is challenging due to GPU memory constraints, necessitating alternate loss functions that can create a similar geometry with smaller batch sizes (e.g., $B=1$).
\end{itemize}

Our work addreses these three challenges.

First, to \textbf{evaluate} long-context retrieval performance, we construct  LoCoV1 (\autoref{fig:main_figure}), a novel benchmark consisting of 12 tasks drawn from law, medicine, science, finance, corporate governance, government reports, and more. LoCoV1 tasks are drawn from real-world datasets spanning diverse domains, including Tau Scrolls, QASPER, LongBench, and the Legal Case Reports corpus \cite{shaham2022scrolls, dasigi2021dataset, bai2023longbench, misc_legal_case_reports_239}. Unlike previous benchmarks, performance on LoCoV1 requires long-context reasoning, and naive truncation and chunking baselines perform poorly (\autoref{tab:m2_on_loco_complete_results}).  

Next, we present the \textbf{M2-BERT retrieval encoder}, an 80M parameter long-context retriever based on the Monarch Mixer architecture \cite{fu2023monarch} and capable of processing up to 32K-length sequences, generating embeddings substantially faster than Transformer-based encoders. 
To \textbf{pretrain} M2-BERT to reason over both short and long contexts, the initial model is pretrained on a mixture of short and long text sequences from C4, Wikipedia, and BookCorpus \cite{2019t5, wikidump, Zhu_2015_ICCV}. Building beyond prior pretraining frameworks for M2-BERT, the long-context versions of this model are also warm-started from shorter-context checkpoints to ensure convergence.

To \textbf{finetune} M2-BERT for retrieval, we explore two alternative strategies that aim to achieve the same embedding geometry as contrastive loss, but are batch-size independent.
    First, we explored prototype loss (PL) \cite{li2021prototypical}, but found weak performance for downstream retrieval.
    Instead, we turned to orthogonal projection loss (OPL) \cite{ranasinghe2021orthogonal},
    which allowed more degrees of freedom for aligning the embeddings of query-passage pairs.
    Furthermore, unlike the common MNRL, OPL optimizes the distance between a query and any relevant/irrelevant document while only requiring a batch size of $B=1$ (\autoref{fig:main_figure}). This allows for finetuning with single-sample batches that fit in memory.

\textbf{Results} Experiments comparing the M2-BERT retrieval encoder to competitive baselines illustrate both performance and efficiency advantages (\autoref{fig:main_figure}). 
In a dense retriever setting, the M2-BERT retrieval encoder substantially outperforms models 5x to 90x its size, beating zero-shot E5-Mistral (7.11B) by 23.3 points and fine-tuned BGE-Large (335M) by 29.9 points on average for LoCoV1 \cite{wang2023improving, bge_embedding}.
M2-BERT also outperforms other retrieval approaches, such as ColBERTv2 \cite{santhanam-etal-2022-colbertv2}, a retrieval model that trades off additional compute at inference time for higher quality, and BM25 \cite{jones2000probabilistic}, a bag-of-words retrieval function that scales easily to longer contexts.
With only 80 million trainable parameters, M2-BERT beats several popular API services, such as OpenAI's \textit{text-embedding-ada-002}, Voyager's \textit{voyage-01}, and Cohere's \textit{embed-english-v3.0} by 35.4 points, averaged across the LoCoV1 datasets. M2-BERT is also 3 to 676$\times$ more efficient at embedding generation than the next state-of-the-art Transformer-based model (E5-Mistral) while also being pretrained on substantially less data.
We provide model checkpoints for the 128, 2048, 8192, and 32768-maximum sequence length versions of the M2-BERT retrieval encoder.

Early open-source previews of the M2-BERT retrieval encoder have been adopted in industry, and an early preview version of LoCo (LoCoV0) is already being used to evaluate new long-context retrieval encoders \cite{günther2024jina, nussbaum2024nomic} (\autoref{tab:loco_v0_results}).

Overall, our work makes the following contributions: (1) the M2-BERT retrieval encoder, a state-of-the-art retriever and the first retriever utilizing a state-space architecture, (2) the long-context (LoCoV1) retrieval benchmark for evaluating and comparing approaches to long-context retrieval, (3) a pretraining and fine-tuning framework for training new M2-BERT retrieval encoders, and (4) an experimental study of the M2-BERT retrieval encoder that illustrates its strengths and weaknesses on long-context tasks.

\section{Related Work}

We overview existing retrieval benchmarks and contrast them with LoCoV1.
We also describe existing state-of-the-art approaches for retrieval models and compare them to M2-BERT.

\parastart{Retrieval Benchmarks} There are a variety of existing retrieval benchmarks for guiding embedding development, such as BEIR, TREC, NaturalQuestions (NQ), SQuAD, and LoTTE~\cite{thakur2021beir, voorhees2005trec, kwiatkowski2019natural, rajpurkar2018know, santhanam-etal-2022-colbertv2}
While these datasets cover a wide breadth of domains, none of them reliably gauge long-context handling during retrieval.
The Tau Scrolls datasets \cite{shaham2022scrolls} seek to gauge long-context handling in language models but it focuses on other knowledge-intensive tasks, such as summarization, fact verification, and natural language inference. 
With the Long-Context (LoCo) Benchmark (V1), we seek to accurately gauge long-context handling in retrieval encoders. 
We selected datasets for which increases to a model's maximum input context will substantially improve retrieval accuracy.

\parastart{Embedding Models for Retrieval} Embedding models are frequently utilized in machine learning pipelines during retrieval.
Many neural embedding models utilize an encoder-only Transformer architecture \cite{vaswani2023attention} that is fine-tuned to maximize cosine similarity between queries and their relevant passages \cite{Reimers2019SentenceBERTSE, leszczynski2022tabi, chen2022perfectly}.
Alternative neural retrieval approaches have emerged to further boost retrieval accuracy while minimizing growing training time, inference time, and memory utilization: examples include dense passage retrieval (DPR) \cite{karpukhin-etal-2020-dense}, late-interaction techniques with ColBERTv2 \cite{santhanam-etal-2022-colbertv2}, and sparse lexical representations with SPLADEv2 \cite{10.1145/3477495.3531833}.
However, new embeddings models based on the generative pretrained transformer (GPT) architecture, such as SGPT, BGE, and E5-Mistral, have reached state-of-the-art accuracy on the BEIR retrieval benchmark \cite{thakur2021beir}, leading to higher quality embedding representations that increase domain generalization \cite{muennighoff2022sgpt, zhang2023retrieve, wang2023improving}. 

In ML pipelines, researchers and practitioners have sought to avoid longer contexts by simply chunking the passages into smaller inputs and averaging the embeddings \cite{lewis2021retrievalaugmented}.
However, for long-context benchmarks, we found that the M2-BERT retrieval encoder outperforms existing models, both when they truncate the input context and when they employ chunking strategies (\autoref{tab:loco_table} and \autoref{tab:m2_on_loco_complete_results}).
This finding suggests that there is indeed a benefit to being able to retrieve over full documents, rather than employing chunking strategies.

\section{LoCoV1 Retrieval Benchmark}

We first motivate the need for retrieval benchmarks which \textit{require} long-context reasoning. We find that on existing benchmark datasets, context length does not correlate with performance, and short-context models yield near state-of-the-art performance. We then describe LoCoV1, which consists of retrieval tasks with long documents. We empirically illustrate that on LoCoV1, performance is more correlated with context length, suggesting that LoCoV1 better measures long-context retrieval abilities.

\begin{table}[]
\centering
\small
\begin{tabular}{ccccc}
\toprule
\textbf{Dataset}      & \textbf{Model}                                                & \textbf{\begin{tabular}[c]{@{}c@{}}Max. Seq.\\ Length\end{tabular}} & \textbf{Score} & \textbf{\begin{tabular}[c]{@{}c@{}}$\Delta$ vs.\\ SOTA\end{tabular}} \\ \midrule
\multirow{3}{*}{\rotatebox[origin=c]{90}{BEIR}} & E5-Mistral                                                    & 4096                                                                & 56.9           & 0.0                                                                  \\
                      & OpenAI Ada                                                    & 8192                                                                & 53.3           & -3.6                                                                 \\
                      & \begin{tabular}[c]{@{}c@{}}BGE-Large\end{tabular} & 512                                                                 & 54.3           & -2.6                                                                 \\ \midrule
\multirow{3}{*}{\rotatebox[origin=c]{90}{LoCo}} & E5-Mistral                                                    & 4096                                                                & 71.4           & -19.7                                                                \\
                      & OpenAI Ada                                                    & 2048*                                                               & 63.4           & -27.7                                                                \\
                      & \begin{tabular}[c]{@{}c@{}}BGE-Large\end{tabular} & 512                                                                 & 54.8           & -36.3   \\ \bottomrule                                                            
\end{tabular}
\caption{BEIR vs. LoCoV1 on Truncation-Based Approaches. We truncate Ada embeddings at 2048 tokens since it is scores higher than truncating at the 8192 max length. SOTA on BEIR is E5-Mistral while SOTA on LoCoV1 is M2-BERT-32k.}
\label{tab:beir_vs_loco_on_truncation}
\end{table}

\parastart{Existing Benchmarks}  We explore whether existing retrieval benchmark datasets adequately capture regimes in which long-context reasoning is \textit{essential} for high performance. We examine BEIR  \cite{thakur2021beir} within the MTEB leaderboard \cite{muennighoff2022mteb}, a popular retrieval benchmark consisting of 17 tasks spanning different domains, query formats, document formats, and query-to-document ratios. In Table \ref{tab:beir_vs_loco_on_truncation}, we compare performance for three high-scoring models with different sequence lengths (using trunction): \textit{E5-Mistral} (4096 tokens), \textit{OpenAI Ada} (8192), and \textit{BGE-Large} (512). First, we observe that the best performing retrieval model, \textit{E5-Mistral}, is only 2.6 accuracy points, on average, ahead of \textit{BGE-Large-en-v1.5}, despite handling $8\times$ longer input sequence length (e.g. 4096 vs. 512). Second, we observe that for most BEIR tasks, the longest documents are only several thousand tokens (Figure \ref{fig:beir_violins}). 
Qualitatively, we note that many BEIR examples have overlap between the query and the beginning of the document (\autoref{tab:beir_examples}).
Overall, these findings suggest that existing benchmark tasks do not effectively capture real-world scenarios where long context retrieval is essential for the downstream ML pipeline (e.g. long context documentation in medicine, law, finance, and more).

\parastart{LoCoV1}  Through the LoCoV1 benchmark, we hoped to encompass a new set of naturalistic, domain-specific retrieval tasks that reflect real-world use cases for long-context queries and documents. LoCoV1 draws from several existing long-context benchmarks, including Tau Scrolls \cite{shaham2022scrolls}, LongBench \cite{bai2023longbench}, and QASPER \cite{dasigi2021dataset}, as well as several domain-specific datasets not originally intended for retrieval, like CourtListener, the Australian Legal Court Reports dataset \cite{misc_legal_case_reports_239}, and the StackOverflow forum.
(details about each task can be found in \autoref{tab:loco_table}). Each dataset was selected for both \textbf{a)} the longer, more complex formatting of its queries and documents as well as \textbf{b)} its ability to gauge long-context handling by containing relevant information throughout its queries and documents. Violin plots depicting document lengths for each of the LoCoV1 tasks can be found in Figure \ref{fig:loco_violins}.

In Table \ref{tab:beir_vs_loco_on_truncation}, we provide results for the same three encoders on LoCoV1. In contrast to BEIR, we find that the relative performance of each model correlates with its sequence length. Additional experiments on LoCoV1 are described in \ref{sec:experiments}.

\section{M2-BERT Retrieval Encoder}\label{sec:M2_description}
Motivated by the need for longer-sequence reasoning on LoCoV1,
we describe (1) the architecture for the M2-BERT retrieval encoder, (2) how the base model is pretrained to reason over both short and long sequences, and (3) how finetuning is performed while respecting GPU memory limits. For notational clarity, we let $\seqlen$ denote maximum sequence length.

\subsection{Architecture}
The M2-BERT retrieval encoder relies on the Monarch Mixer (M2) architecture, a BERT-like model that utilizes Monarch matrices for language modeling. 
Monarch Mixer is part of a new class of architectures called \textit{state-space models} (SSMs), which include S4, Mamba, and BiGS \cite{gu2022efficiently, gu2023mamba, wang2022pretraining}. 
Unlike regular BERT and long-context Transformer-based encoder like LongFormer \cite{Beltagy2020Longformer}, M2-BERT can handle longer input contexts by leveraging Monarch matrices as a subquadratic primitive along both input sequence length and model dimension.
While new Transformer-based models capable of encoding 8k tokens have emerged \cite{günther2023jina}, the M2-BERT encoders can handle up to 32k input tokens, undergo fine-tuning substantially faster than attention-based models, run inference 3 to 676x more rapidly (\autoref{tab:m2_bert_efficiency}), and still achieve state-of-the-art on long context retrieval tasks.

\subsection{Pretraining}
Retrieval encoders frequently rely on model backbones which have already been pretrained on corpora from the relevant language \cite{Reimers2019SentenceBERTSE, santhanam-etal-2022-colbertv2, karpukhin-etal-2020-dense, 10.1145/3477495.3531833}. This equips the model with the capacity to understand and reason over text sequences, and enables high performance even when the retrieval-specific finetuning dataset is small \cite{saad-falcon-etal-2023-udapdr}. The difficulty with using the M2 architecture for our encoder is that previous work has only (1) studied pretraining M2 for sequence lengths up to 128 tokens, and (2) studied pretraining in regimes where downstream tasks consisted of sequences mostly uniform in length (e.g., short GLUE tasks). In contrast, the long-context retrieval setting requires that the base model be capable of understanding both short sequences (for queries) and long sequences (for documents).

The first technical challenge is designing a pretraining dataset over which the masked language modeling (MLM) objective enables the model to learn both short and long sequences. Experimentally, we find that training with only short or only long sequences is insufficient, and that instead the pretraining data must contain a mixture of both short and long context samples (see \ref{sec:ablation} for comparisons to alternative strategies). For the source of these samples, we rely on three high quality datasets routinely used for pretraining: C4 \cite{2019t5}, Wikipedia \cite{wikidump}, and BookCorpus \cite{Zhu_2015_ICCV}. For our short context examples, we include variable length passages from our three training corpora, which can range from 10 tokens to our maximum input sequence length of 128, 2048, 8192, or 32768 tokens, depending on the M2-BERT model. For our long context examples, we concatenate multiple successive training examples together to generate sequences that reach our maximum input sequence length.

\begin{table}[]
\small 
\centering
\begin{tabular}{cccc} 
      \toprule
      \textbf{Length Type}   &  \textbf{C4} & \textbf{Wikipedia} & \textbf{BookCorpus}\\ \midrule
       Variable   & 10\% & 10\% & 10\% \\
       Maximum   & 24\% &  23\% & 23\% \\ \bottomrule
    \end{tabular}
    \caption{Pretraining dataset proportions based on text source and sequence length type of the training examples.}
    \label{tab:pretraining_mix}
\end{table}

The second technical challenge is ensuring pretraining convergence when the maximum sequence length is greater. We find that traditional initialization with random weights is sufficient when $\seqlen \in \{128, 2\text{k}, 8\text{k}\}$. For $\seqlen = 32\text{k}$ however, we find that models initialized with random weights do not converge to sufficient MLM accuracies within a reasonable amount of time. Therefore to accelerate training convergence for this model, we warm start with the weights of a pretrained 8k checkpoint, and initialize the 32k positional embeddings with the initial 8k positional embeddings by extending them through replication across the newly initialized weights. Under this strategy, the 32K model converges.

\subsection{Fine-tuning}
To adapt a pretrained model for a specific retrieval task (e.g., identifying the relevant legal case given a description), it is common practice to finetune that model on a collection of representative queries and documents \cite{Reimers2019SentenceBERTSE, santhanam-etal-2022-colbertv2, karpukhin-etal-2020-dense, saad-falcon-etal-2023-udapdr}.

\textbf{MNRL} A popular approach is to finetune the base model using a contrastive learning loss called multiple negatives ranking loss \cite{henderson2017efficient}, which encourages the model to learn embeddings of queries and documents for which the cosine similarity of relevant query-document pairs is high, and irrelevant query-document pairs is low. It requires a dataset of query and relevant document pairs ($\{(q_i, d_i)\}_{i=1}^n$. For a query $q_i$, MNRL samples $k$ random documents from $\{d_j\}_{j=1, j\neq i}^n$ as ``negative'' passages, and generates a ``prediction'' for $q_i$ against $d_i$ and the $k$ distractors by computing pairwise cosine similarities (e.g. $\textit{PCS}$). CrossEntropyLoss (e.g. $\textit{CE}$) is applied to these predictions, treating the $k$ distractors as the negative class and $d_i$ as the positive class.  For a given query $q_k$, we compute MNRL as:
\begin{align*}
    \textit{MNRL}(\{q_k, d_i\}_{i=1}^n) &= \textit{CE}(\text{Scores}, \text{Labels})\\
    \text{Scores} &= [\textit{PCS}(q_k, d_i)_{i=1}^n]\\
    \text{Labels} &=[1,..., n]
\end{align*}
MNRL is closely related with constrastive loss, and induces an embedding geometry of \textit{alignment} between query-document pairs, and \textit{uniformity} of document embeddings across the hypersphere \cite{wang2022understanding, chen2022perfectly, leszczynski2022tabi, fu2022details}. This loss function requires large batch sizes for quality.

In MNRL, a single query and all $k+1$ documents must fit within a single batch. In the long-context regime, GPU memory requirements thus force a tradeoff between $k$ and $\seqlen$. When $\seqlen$ is small (e.g., 128 tokens), $k$ can be large and still fit in GPU memory (e.g., $k = 128$). When $\seqlen$ is large however (e.g., 32k tokens), the memory footprint of a single document is larger, and $k$ must be considerably smaller (e.g., $k = 2$). The technical challenge is that MNRL only works well for large $k$~\cite{henderson2017efficient}, and thus, suboptimal for long sequences (see Sec. \ref{sec:ablation}).

\textbf{Prototype Loss} In our work, we seek a method to achieve the same embedding geometry as MNRL, but in a batch-independent way. 
One approach is \textit{prototype loss} \cite{li2021prototypical}, which uses a target model's embeddings to guide the contrastive learning of a student model.
By leveraging the learned embeddings of a model trained with MNRL (e.g. M2-BERT-128), we may be able to rapidly fine-tune a long-context embedding model that is limited to a much smaller batch size (e.g. M2-BERT-32k).
Given query $q_k$, passage $p_k$, target embedding model $\textit{TM}$, and student embedding model $\textit{SM}$, we calculate prototype loss (PL) as: 
\begin{align*}\label{eq:pl}
    PL(\{q_k, p_k\}) &= \text{Query Loss} + \text{Passage Loss} \\
    \text{Query Loss} &= \textit{PCS}(\textit{TM}(q_k), \textit{SM}(q_k)) \\
    \text{Passage Loss} &=\textit{PCS}(\textit{TM}(p_k), \textit{SM}(p_k))
\end{align*}

Even if our M2-BERT-32k model successfully learns the embeddings from the M2-BERT-128 model, it still requires further fine-tuning from the starting M2-BERT-128 representations to develop robust embeddings for 32k context length.
After using prototype loss to fine-tune our M2-BERT-32k with the fine-tuned M2-BERT-128 model as the target embeddings (\autoref{tab:opl_v_mnrl}), we find that the M2-BERT-128 embeddings are not the ideal starting weights for further fine-tuning of M2-BERT-32k; the learned representations at 128 context length are substantially different than the learned representation at 32k context length \autoref{sec:m2_training_finetuning_eval}).

\textbf{Orthogonal Projection Loss} To overcome these challenges, we instead finetune our M2-BERT base model using \textit{orthogonal projection loss} (OPL) \cite{ranasinghe2021orthogonal}. 
Unlike MNRL, OPL is compatible with single-sample batches by using Mean Squared Error (e.g. $\textit{MSE}$).  
Unlike prototype loss, OPL does not require a teacher model for embeddings.
Given a query $q_k$ and passage $p_k$, we calculate OPL as: 
\begin{align*}
    OPL(\{q_k, p_k\}) &= \textit{MSE}(\text{Score}, \text{Label})\\
    \text{Score} &= \textit{PCS}(q_k, p_k)\\
    \text{Label} &= 1.0 \text{~for positives, $0.0$ for negatives}
\end{align*}

Intuitively, OPL finetunes the model to encourage embeddings for positive query-document pairs to be aligned with each other, and for negative query-document pairs to be orthogonal to each other. Because OPL operates on a single query-document pair, it performs well on single-sample batches, and is thus ideal for our long-context setting. Similar to MNRL, we sample negative documents for query $q_i$ from $\{d_j\}_{j=1, j\neq i}^n$.
Lastly, we note that while OPL proves effective for fine-tuning our M2-BERT encoder (Section \ref{sec:M2_quality}), OPL is just one choice of loss function; other functions with similar properties may be useful.

\section{Experiments}\label{sec:experiments}
Our experimental evaluations focus on three questions: (1) How does the M2-BERT retriever compare to existing baselines (in terms of quality and efficiency) for retrieval over both long context and short context documents? (2) How necessary are the pretraining and finetuning approaches proposed in \autoref{sec:M2_description}, and how do they compare to standard retriever pretraining/finetuning methods? (3) Can the representations learned by the fine-tuned M2-BERT models be used for non-retrieval tasks, like data visualization or clustering-based classification?

\subsection{Comparing M2-BERT to Existing Retriever Models}
\label{sec:M2_quality}

\begin{table}[]
\centering
\small
\setlength{\tabcolsep}{2.2pt}
\begin{tabular}{ccccc}
\toprule
\textbf{Model}                               & \begin{tabular}[c]{@{}c@{}}\textbf{Param.}\\ \textbf{Count}\end{tabular} & \begin{tabular}[c]{@{}c@{}}\textbf{Max. Seq.}\\ \textbf{Length}\end{tabular} & \begin{tabular}[c]{@{}c@{}}\textbf{LoCoV1}\\ \textbf{Score}\end{tabular} & \begin{tabular}[c]{@{}c@{}}\textbf{LoCoV1 Score}\\ \textbf{w. Chunks}\end{tabular} \\ \midrule
\begin{tabular}[c]{@{}c@{}}BGE-Large\\ Zeroshot\end{tabular}     & 335M                                   & 512                                         & 56.5 & 54.8                                  \\ \midrule
\begin{tabular}[c]{@{}c@{}}BGE-Large\\ Finetuned\end{tabular}    & 335M                                   & 512                                         & 64.8  & 61.6           \\ \midrule
E5-Mistral & 7.11B                                  & 4096                                        & 71.4  & 70.3                                \\ \midrule
BM25 & N/A                                  & N/A                                        & 79.9  & N/A                                \\ \midrule
\begin{tabular}[c]{@{}c@{}}Jina Embeds.\end{tabular}           & 137M                                   & 8192                                        & 67.2 &  19.2                                 \\ \midrule
\begin{tabular}[c]{@{}c@{}}OpenAI Ada\end{tabular}             & N/A                                    & 8192                                        & 63.2  &  63.4
\\ \midrule
\begin{tabular}[c]{@{}c@{}}ColBERTv2\end{tabular}            & 110M                                    & 512                                         & 53.6  & N/A
\\ \midrule
\begin{tabular}[c]{@{}c@{}}M2-BERT-128\end{tabular}            & 80M                                    & 128                                         & 70.3  & N/A            \\ \midrule
\begin{tabular}[c]{@{}c@{}}M2-BERT-2k\end{tabular}           & 80M                                    & 2048                                        & 82.3  & N/A            \\ \midrule
\begin{tabular}[c]{@{}c@{}}M2-BERT-8k\end{tabular}           & 80M                                    & 8192                                        & 86.9  & N/A            \\ \midrule
\begin{tabular}[c]{@{}c@{}}M2-BERT-32k\end{tabular}          & 80M                                    & 32768                                       & \textbf{94.7} & N/A   \\ \bottomrule         
\end{tabular}
\caption{M2-BERT Retrieval Encoder and Baseline Model Performances on LoCoV1.}
\label{tab:m2_on_loco_abridged}
\end{table}

We begin by evaluating the M2-BERT retriever's performance relative to existing competitive retriever methods. We choose five of the best performing models from BEIR. These are: BGE-Large-en-v1.5 \cite{bge_embedding}, E5-Mistral \cite{wang2023improving}, Jina Embeddings \cite{günther2023jina}, OpenAI Ada embeddings (\textit{text-embedding-ada-002}), and ColBERTv2 \cite{santhanam-etal-2022-colbertv2}.
The Appendix reports additional models that we evaluated but that have worse performance.

The baseline models have maximum sequence lengths shorter than some documents in LoCoV1. We therefore study two approaches for generating embeddings. The first approach truncates each document to the length of the model's maximum sequence length, while the second approach segments the document into chunks (each the size of the model's maximum sequence length) and computes a document embedding as the average of chunk embeddings. All M2-BERT models are evaluated with the LoCoV1 and BEIR retrieval benchmarks.

We use nDCG@10~\cite{wang2013theoretical} as the quality metric for LoCoV1.
nDCG@10 measures the ranking quality of information retrieval systems, accounting for both the position and quality of the items in the retrieved sequence. We evaluate efficiency by calculating the time it takes to embed 32k document tokens, on average, whether that is through one single embedding or multiple chunked embeddings. 
Appendix \ref{sec:efficiency_details} provides additional information.

\parastart{LoCoV1} \autoref{tab:m2_on_loco_abridged} compares averaged nDCG scores for all methods on the LoCoV1 benchmark (\autoref{tab:m2_on_loco_complete_results} provides results by task). 
Performance improvements are significant --- we found that M2-BERT-32k outperformed the next best baseline approach (\textit{BM25}) by an average of 14.8 points, 
the
next best truncation-baseline approach (\textit{E5-Mistral}) by an average of 23.3 points, and 
the next best chunked-baseline approach (\textit{E5-Mistral}) by an average of 24.4 points. On a per-task level, M2-BERT-32k outperforms all baseline methods on 7 of 12 tasks, and all Transformer-based methods on 10 of 12 tasks. %

We also observe that retrieval accuracy increased as we incrementally scaled maximum sequence length of the M2-BERT retrieval encoder for each of our models. The overall performance improvement for going from a sequence length of 128 tokens to 32k tokens is approximately 21.0 points (average). In contrast, alternate retrieval strategies---like chunking---appeared to barely improve other base retrieval models, and sometimes even worsen them.  Overall, our findings demonstrate that standard retrieval approaches, whether it is truncation or chunking with embedding averaging, are not sufficient for handling long-context documents in retrieval, and that M2-BERT outperforms baseline models while being substantially smaller. 

\parastart{BEIR} We study whether long-context M2-BERT retrieval models sacrifice short-context performance by evaluating on the BEIR benchmark (\autoref{tab:m2_vs_sbert_on_beir}).
We compare to SentenceBERT, a language model of comparable size with a similar pretraining ensemble and identical fine-tuning process for BEIR (e.g. fine-tuning on the MS MARCO retrieval dataset \cite{bajaj2018ms}).
We do not compare against other high-performing models on BEIR, such as BGE-Large \cite{bge_embedding} and E5-Mistral \cite{wang2023improving}, since they are substantially larger than M2-BERT and use significantly more datasets for both pretraining and embedding fine-tuning, making it difficult to compare training and architecture selections directly.
We find that M2-BERT-128 approximately matches SentenceBERT performance, averaging 1.3 nDCG@10 points lower, and performs better than SentenceBERT on some of the longer context classification datasets (e.g. AmazonPolarityClassification and AmazonReviewsClassification).

\begin{table}[]
\centering
\small
\setlength{\tabcolsep}{2.3pt}
\begin{tabular}{cccccc}
\toprule
\textbf{Model}                                               & \textbf{\begin{tabular}[c]{@{}c@{}}Max. Seq.\\ Length\end{tabular}} & \textbf{\begin{tabular}[c]{@{}c@{}}Param.\\ Count\end{tabular}} & \textbf{\begin{tabular}[c]{@{}c@{}}BEIR\\ Score\end{tabular}} & \textbf{\begin{tabular}[c]{@{}c@{}}$\Delta$ \\ Params\end{tabular}} & \textbf{\begin{tabular}[c]{@{}c@{}}$\Delta$\\ BEIR\\ Score\end{tabular}} \\ \midrule
SentenceBERT                                                 & 512                                                                 & 110M                                                            & 40.0                                                         & 0\%                                                                 & 0                                                                      \\ \midrule
\begin{tabular}[c]{@{}c@{}}M2-BERT-128\end{tabular} & 128                                                                 & 80M                                                             & 38.7                                                         & -27\%                                                               & -1.3  \\ \bottomrule                                                                 
\end{tabular}
\label{tab:m2_vs_sbert_on_beir}
\caption{M2-BERT vs. SentenceBERT on BEIR.}
\end{table}

\parastart{Computational Efficiency} We compare M2-BERT to baseline methods in terms of throughput, i.e., the time it takes to both tokenize and embed the entirety of an $X$ token document (\autoref{tab:m2_bert_efficiency}) on A100. For models that cannot tokenize and embed the $X$ tokens all at once, we create separate embeddings for $Y$ token chunks, where $Y$ is the maximum sequence length of the model. We find that M2-BERT-32K provides the greatest throughput, producing an embedding $3.13\times$ more efficiently for a 512 token document and $676\times$ more efficiently for a 32768 token document relative to the next best state-of-the-art model, \textit{E5-Mistral}.

\begin{table}[h!]
\centering
\small
\setlength{\tabcolsep}{2.2pt}
\begin{tabular}{cccccc}
\toprule
\multicolumn{1}{l}{}                                             & \multicolumn{1}{l}{}                                                & \multicolumn{4}{c}{\textbf{Time to Encode $X$ Tokens}}           \\ \midrule
\textbf{Models}                                                  & \textbf{\begin{tabular}[c]{@{}c@{}}Max. Seq.\\ Length\end{tabular}} & \textbf{128}              & \textbf{2048}             & \textbf{8192}            & \textbf{32768}           \\
\midrule
BGE-Large                                                        & 512                                                                 & 0.015                     & 0.029                     & 0.12                     & 0.49                     \\ \midrule
E5-Mistral & 4096                                                                & 0.029                     & 0.11                      & 1.2                      & 4.8                      \\ \midrule
Jina Embeds.                                                     & 8192                                                                & 0.0070                    & 0.0070                    & 0.0070                   & 0.028                    \\ \midrule
M2-BERT-128                                                          & 128                                                                 & 0.028                     & 0.057                     & 0.12                     & 0.46                     \\ \midrule
M2-BERT-2k                                                          & 2048                                                                & 0.0071                    & 0.0071                    & 0.028                    & 0.057                    \\ \midrule
M2-BERT-8k                                                          & 8192                                                                & 0.0072                    & 0.0072                    & 0.0072                   & 0.028                    \\ \midrule
M2-BERT-32k                                                          & 32768                                                               & 0.0071                    & 0.0071                    & 0.0071                   & 0.0071                   \\ \midrule
\multicolumn{2}{c}{\textbf{$\Delta$ 32k Speed vs. Mistral}}                                         & 3.13x & 14.9x & 169x & 676x \\ \bottomrule
\end{tabular}
\caption{M2-BERT Efficiency Comparison to Baseline Models.}
\label{tab:m2_bert_efficiency}
\end{table}

\parastart{Needle-in-the-Haystack Synthetic}
We perform a more detailed analysis of the M2-BERT retriever's ability to encode long contexts and capture relevant information, despite surrounding irrelevant context, by using a synthetic modeled off ``needle-in-the-haystack" tasks that have been used in other studies of longer context tasks \cite{liu2023lost}.
For our version, we adapted the Natural Questions (NQ) benchmark \cite{kwiatkowski2019natural}, which contains contains query-relevant passage pairs (derived from Google and Wikipedia). We use the original queries provided by the NQ benchmark, but modified the passages by adding 39 "distractor" passages to each relevant passage in the answer set. These distractor passages are selected by randomly sampling other Wikipedia passages. We study how the location of the relevant passage (e.g., appearing first vs. appearing seventh amongst all the passages) impacted retrieval performance. Since we have 40 passages total, there are exactly 40 different positions to place the relevant passage within the sequence.

\begin{figure}[t]
   \centering
   \includegraphics[width=0.7\linewidth]{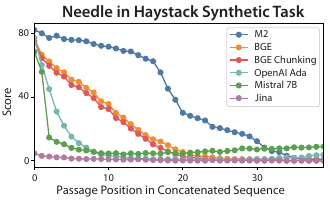}
   \caption{M2-BERT and Baseline Model Performance on Needle-in-the-Haystack Synthetic Task.}
   \label{fig:synthetic_task}
\end{figure}

We compare M2-BERT-32k to the best-performing baselines: \textit{BGE-Large-en-v1.5}, \textit{E5-Mistral} and Jina Embeddings (\autoref{fig:synthetic_task}). We observe a relationship between the position of the relevant passage and the relative performance improvement of M2-BERT-32k. When the relevant passage is closer to the start of the concatenated sequence, the baseline models perform almost as well as M2-BERT-32k. However, as the relevant passage moves to the end of the concatenated sequence, the performances of the baseline models substantially drops since the models cannot see the relevant passage within the total sequence, due to their shorter maximum sequence lengths (see \autoref{tab:m2_on_synth_task_complete_results} for complete results).

\subsection{Ablation of Pretraining and Finetuning}\label{sec:ablation}

\begin{table}[]
\small 
\centering
\begin{tabular}{cccc}
\toprule
\textbf{Model} & \textbf{\begin{tabular}[c]{@{}c@{}}Max. Seq.\\ Length\end{tabular}} & \textbf{\begin{tabular}[c]{@{}c@{}}Training\\ Selection\end{tabular}} & \textbf{\begin{tabular}[c]{@{}c@{}}LoCoV1\\ Score\end{tabular}} \\ \midrule
M2-BERT        & 2048                                                                & Short Examples                                                        &  37.2                                                                     \\ \midrule
M2-BERT        & 2048                                                                & Long Examples                                                         &     44.9                                                                 \\ \midrule
M2-BERT        & 2048                                                                & Mixed Examples                                                        &   \textbf{55.4}   \\ \bottomrule                                                                
\end{tabular}
\caption{M2-BERT Training Example Selection for Pretraining.}
\label{tab:m2_pretraining_examples}
\end{table}

\begin{table}[]
\small 
\centering
\begin{tabular}{cccc}
\toprule
\textbf{Model} & \textbf{\begin{tabular}[c]{@{}c@{}}Max. Seq.\\ Length\end{tabular}} & \textbf{\begin{tabular}[c]{@{}c@{}}Checkpoint\\ Selection\end{tabular}} & \textbf{\begin{tabular}[c]{@{}c@{}}MLM\\ Accuracy\end{tabular}} \\ \midrule
M2-BERT-32k        & 32768                                                                & Warm-Start                                                              & \textbf{33.9}                                                                \\ \midrule
M2-BERT-32k        & 32768                                                                & Cold-Start                                                              & 4.8                                                                \\ \bottomrule                                                             
\end{tabular}
\caption{Warm vs. Cold Start for M2-BERT-32768 Pretraining - MLM Train Accuracy after 6,000 Training Steps.}
\label{tab:m2_bert_pretraining_checkpoint}
\end{table}

Section \ref{sec:M2_description} presents two design choices for the M2-BERT retriever---pretraining data mixture and finetuning loss objective. This subsection evaluates those choices in comparison to alternative pretraining and finetuning approaches. 

\parastart{Pretraining} In Section \ref{sec:M2_description}, we describe selecting a pretraining mixture for the M2-BERT base model consisting of both short and long sequences. We compare this to two alternate pretraining regimes: (1) solely using short training examples, and (2) solely using long training examples (Table \ref{tab:m2_pretraining_examples}). For each regime, we pretrain the M2-BERT-2048 architecture to 5,000 training steps before further fine-tuning on the LoCoV1 dataset but with a limited number of negatives (e.g. 8 negative passages per query-positive passage pair). 
We observe that the model trained on the mixed short/long sequence dataset performs best, beating solely long sequence pretraining by 10.5 points on average.

We also illustrate the necessity of initializing M2-BERT-32k with the weights of a M2-BERT-8k checkpoint (\autoref{tab:m2_bert_pretraining_checkpoint} and \autoref{fig:warm_vs_cold_checkpoints_graph}). Compared to random initialization, we find that the version with warm-starting converges dramatically faster, successfully completing pretraining in the same number of steps as our other M2-BERT encoders.

\parastart{Finetuning} Section \ref{sec:M2_description} describes how GPU memory constraints limit the  training batch size for longer context M2-BERTs, necessitating the use of OPL loss function, which can function with single-sample batch sizes. We illustrate the batch size-performance tradeoff incurred by the traditionally used MNRL loss function by comparing (1) OPL trained with batch size 1, to (2) MNRL trained with the maximum batch size possible on an A100 GPU (\autoref{tab:opl_v_mnrl}).
For fine-tuning M2-BERT-32k, we find that OPL improved average nDCG@10 on LoCoV1 by 29.4\% compared to MNRL.

\begin{table}[t]
\centering
\small
\begin{tabular}{ccccc}
\toprule
\textbf{Model}       & \textbf{\begin{tabular}[c]{@{}c@{}}Loss \\ Function\end{tabular}} & \textbf{\begin{tabular}[c]{@{}c@{}}Batch\\ Size\end{tabular}} & \textbf{\begin{tabular}[c]{@{}c@{}}LoCoV1\\ Score\end{tabular}} & \textbf{\begin{tabular}[c]{@{}c@{}}$\Delta$\\ Scores\end{tabular}} \\ \midrule
M2-BERT-32k & MNRL                                                     & 2                                                    & 70.4                                                      & 0                                                       \\
\midrule
M2-BERT-32k & PL                                                      & 2  & 63.2 & -7.2 \\
\midrule
M2-BERT-32k & OPL                                                      & 1                                                    & \textbf{94.7}                                                      & 24.3 \\ 
\bottomrule
\end{tabular}
\caption{OPL vs. MNRL for Fine-tuning M2-BERT-32k.}
\label{tab:opl_v_mnrl}
\end{table}
\subsection{Applications of M2-BERT Retrieval Encoders}

Finally, we explore whether the embeddings from the M2-BERT retrieval model are useful for other embedding tasks. 

\begin{figure}[t]
   \centering
   \includegraphics[width=0.7\linewidth]{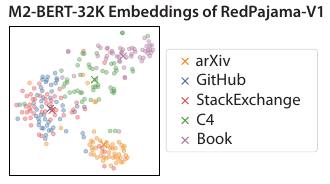}
   \caption{t-SNE Visualization of M2-BERT-32K Embeddings of RedPajama-V1 sample.}
   \label{fig:tsne_visualization}
\end{figure}

\parastart{Zero-shot Clustering} We find our M2-BERT retrieval encoders can be used effectively for zero-shot clustering of textual datasets. Using our M2-BERT-32k model, we take a sample of the RedPajama-v1 dataset \cite{together2023redpajama} and generate embeddings for datapoints from each of the constituent datasets: C4, StackExchange, BookCorpus, ArXiv, and Github. In \autoref{fig:tsne_visualization}, we visualize the M2-BERT embeddings for the sampled datapoints from RedPajama. We find that the datapoints for Github and StackExchange tend to be grouped together, likely due to their overlapping subject terminologies. Additionally, we find limited overlap between C4 and BookCorpus due to some shared subjects between the two constituent datasets. Lastly, the ArXiv datapoints seem mostly isolated by its unique mix of technical topics.

\parastart{MTEB with M2} To further explore the robustness of the M2-BERT embeddings, we test our M2-BERT retrieval encoders on the MTEB benchmark tasks. In \autoref{tab:m2_mteb_abridged}, we compare the zero-shot results of our M2-BERT-128 retrieval encoder to the SentenceBERT baseline for MTEB benchmark, evaluating on only the English datasets, which cover classification, clustering, pair classification, reranking, and semantic textual similarity (STS) (for the expanded results, see Appendix \ref{sec:mteb_benchmark_complete_results}). We found that M2-BERT-128 performed comparably to the SentenceBERT model, scoring 0.2 accuracy points higher than SentenceBERT, on average, despite substantially less pretraining data and 27\% less parameters. We are interested to explore further applications of M2-BERT in both classification and clustering tasks, particularly for long-context tasks.

\begin{table}[]
\small
\centering
\setlength{\tabcolsep}{2.5pt}
\begin{tabular}{cccc}
\toprule
                                     & \textbf{Model}                                                      & SentenceBERT & M2-BERT-128 \\ \midrule
                                     & \textbf{\begin{tabular}[c]{@{}c@{}}Max. Seq.\\ Length\end{tabular}} & 512          & 128         \\ \midrule
\multirow{5}{*}{\rotatebox[origin=c]{90}{\textbf{Tasks}}} & Classification                                                      & 64.5         & 63.4        \\
                                     & Clustering                                                          & 33.7         & 32.5        \\
                                     & \begin{tabular}[c]{@{}c@{}}Pair Classification\end{tabular}       & 90.5         & 90.3        \\
                                     & Reranking                                                           & 50.9         & 51.3        \\
                                     & STS                                                                 & 76.1         & 78.8        \\ \midrule
                                     & \textbf{MTEB Avg. Score}                                            & 63.1         & \textbf{63.3}        \\ \midrule
                                     & \textbf{$\Delta$ Params}                                            & 0\%          & -27\%       \\ \midrule
                                     & \textbf{$\Delta$ MTEB Scores}                                       & 0            & +0.2 \\ \bottomrule       
\end{tabular}
\caption{M2-BERT vs. SentenceBERT on MTEB.}
\label{tab:m2_mteb_abridged}
\end{table}

\section{Conclusion}

In this work, we introduce the \textbf{M2-BERT retrieval encoder}, the first retrieval encoder capable of handling contexts of 32k tokens and the first state-space model retriever.
The Monarch Mixer architecture allows our M2-BERT encoders to scale subquadratically with input context length, capably handling long-context queries and documents despite only having 80M trainable parameters.
To better understand how M2-BERT and other retrieval encoders can handle long-context queries and documents, we also developed the \textbf{LoCoV1} retrieval benchmark, a set of twelve expert-annotated datasets spanning law, medicine, science, screenwriting, finance, and more.
Our M2-BERT retrieval encoders match Transformer-based retrieval encoders on the BEIR benchmark while achieving state-of-the-art performance on LoCoV1, beating the next state-of-the-art retrieval encoder by 19.7 accuracy points while being 3 to 676$\times$ more efficient.
We are excited to continue exploring applications of the M2-BERT encoder architecture, such as classification, clustering, and retrieval-augmented generation (RAG), as well as test other promising fine-tuning approaches, such as cached MNRL \cite{henderson2017efficient, gao-etal-2021-scaling}.
We hope our M2-BERT retrieval encoders and the LoCoV1 benchmark will bolster ML pipelines across application domains.

\section{Acknowledgements}

We thank Silas Alberti, Sabri Eyuboglu, Omar Khattab, Gautam Machiraju, Eric Nguyen, Krista Opsahl-Ong, Christopher Potts, Benjamin Spector, Alyssa Unell, Benjamin Viggiano, Michael Wornow, and Michael Zhang for their constructive feedback during the composition of the paper.
We would also like to thank our collaborators at the Stanford Artificial Intelligence Laboratory (SAIL) and TogetherAI.

We gratefully acknowledge the support of NIH under No. U54EB020405 (Mobilize), NSF under Nos. CCF1763315 (Beyond Sparsity), CCF1563078 (Volume to Velocity), and 1937301 (RTML); US DEVCOM ARL under No. W911NF-21-2-0251 (Interactive Human-AI Teaming); ONR under No. N000141712266 (Unifying Weak Supervision); ONR N00014-20-1-2480: Understanding and Applying Non-Euclidean Geometry in Machine Learning; N000142012275 (NEPTUNE); NXP, Xilinx, LETI-CEA, Intel, IBM, Microsoft, NEC, Toshiba, TSMC, ARM, Hitachi, BASF, Accenture, Ericsson, Qualcomm, Analog Devices, Google Cloud, Salesforce, Total, the HAI-GCP Cloud Credits for Research program,  the Stanford Data Science Initiative (SDSI), Stanford EDGE Fellowship, GEM Fellowship Program, and members of the Stanford DAWN project: Facebook, Google, and VMWare. 
Neel Guha is supported by the Stanford Interdisciplinary Graduate Fellowship and the HAI Graduate Fellowship.
The U.S. Government is authorized to reproduce and distribute reprints for Governmental purposes notwithstanding any copyright notation thereon. Any opinions, findings, and conclusions or recommendations expressed in this material are those of the authors and do not necessarily reflect the views, policies, or endorsements, either expressed or implied, of NIH, ONR, or the U.S. Government.

\bibliography{example_paper}
\bibliographystyle{plain}

\appendix
\onecolumn
\section{Appendix}

\subsection{LoCoV1 Overview}
Table \ref{tab:loco_table} provides an overview of the LoCo benchmark.

\begin{table}[h!]
\tiny
\centering
\setlength{\tabcolsep}{2.5pt}
\begin{tabular}{lllcccccc}
\toprule
\textbf{Dataset}               & \textbf{Source}                                                        & \textbf{Domain}                                                & \multicolumn{1}{c}{\textbf{\begin{tabular}[c]{@{}c@{}}\# of Train \\ Queries\end{tabular}}} & \multicolumn{1}{c}{\textbf{\begin{tabular}[c]{@{}c@{}}\# of Train\\ Documents\end{tabular}}} & \multicolumn{1}{c}{\textbf{\begin{tabular}[c]{@{}c@{}}\# of Test\\ Queries\end{tabular}}} & \multicolumn{1}{c}{\textbf{\begin{tabular}[c]{@{}c@{}}\# of Test\\ Documents\end{tabular}}} & {\textbf{\begin{tabular}[c]{@{}c@{}}Avg. Query \\ Length\end{tabular}}} & {\textbf{\begin{tabular}[c]{@{}c@{}}Avg. Doc. \\ Length\end{tabular}}}\\
\midrule
SummScreenFD                   & Tau Scrolls                                                            & Screenwriting                                                  & 3673                                                                                        & 3673                                                                                         & 338                                                                                       & 338 & 590 & 30,792                                                                                        \\ \midrule
Gov. Report                    & Tau Scrolls                                                            & Government                                                     & 17457                                                                                       & 17457                                                                                        & 972                                                                                       & 972   & 3,871 & 55,280                                                                                      \\ \midrule
QMSUM                          & Tau Scrolls                                                            & \begin{tabular}[c]{@{}l@{}}Corporate\\ Management\end{tabular} & 1257                                                                                        & 1257                                                                                         & 272                                                                                       & 272  & 430 & 58,129                                                                                      \\ \midrule
{\begin{tabular}[l]{@{}l@{}}QASPER\\Title to Full Text\end{tabular}}    & QASPER                                                                 & Science                                                        & 888                                                                                         & 888                                                                                          & 416                                                                                       & 416 & 71 & 22,315                                                                                         \\ \midrule
{\begin{tabular}[l]{@{}l@{}}QASPER\\Abstract to Full Text\end{tabular}} & QASPER                                                                 & Science                                                        & 888                                                                                         & 888                                                                                          & 416                                                                                       & 416 & 931 & 22,315                                                                                         \\ \midrule
MultiFieldQA                   & LongBench                                                              & \begin{tabular}[c]{@{}l@{}}General\\ Domain\end{tabular}       & 120                                                                                         & 120                                                                                          & 30                                                                                        & 30   & 62 & 29,465                                                                                       \\ \midrule
2WikimQA                       & LongBench                                                              & \begin{tabular}[c]{@{}l@{}}General\\ Domain\end{tabular}       & 240                                                                                         & 240                                                                                          & 60                                                                                        & 60 & 69 & 37,867                                                                                         \\ \midrule
Passage Retrieval              & LongBench                                                              & \begin{tabular}[c]{@{}l@{}}General\\ Domain\end{tabular}       & 240                                                                                         & 240                                                                                          & 60                                                                                        & 60  & 840 & 35,814                                                                                        \\ \midrule
CourtListener - Plain Text     & CourtListener                                                          & Law                                                            & 10000                                                                                       & 10000                                                                                        & 2000                                                                                      & 2000  & 146 & 48,190                                                                                      \\ \midrule
CourtListener - HTML           & CourtListener                                                          & Law                                                            & 10000                                                                                       & 10000                                                                                        & 2000                                                                                      & 2000    & 146 & 57,028                                                                                    \\ \midrule
\begin{tabular}[c]{@{}l@{}}Australian Legal\\ Case Report\end{tabular}  & \begin{tabular}[c]{@{}l@{}}Australian Legal\\ Case Report\end{tabular} & Law                                                            & 3094                                                                                        & 3094                                                                                         & 770                                                                                       & 770 & 14,986 & 47,536                                                                                        \\ \midrule
StackOverflow                  & StackOverflow                                                          & Programming                                                    & 1599                                                                                        & 18005                                                                                        & 400                                                                                       & 7741 & 758 & 4,544  \\ \bottomrule                                                                                    
\end{tabular}
\caption{Overview of Long-Context (LoCo) benchmark (V1) and its constituent datasets.}
\label{tab:loco_table}
\end{table}
\clearpage

\subsection{LoCoV1 Query and Document Examples}
\label{sec:loco_examples}

\begin{table}[H]
\centering
\tiny
\setlength{\tabcolsep}{1.0pt}
\begin{tabular}{lll}
\toprule
\textbf{\begin{tabular}[c]{@{}c@{}}LoCoV1\\Dataset\end{tabular}}                                                     & \textbf{Query Example}                                                                                                                                                                                                                                                                                                                                                                                                                                                                                                                                                                                                                                                                 & \textbf{Document Example}                                                                                                                                                                                                                                                                                                                                                                                                                                                                                                      \\ 
\midrule
\begin{tabular}[c]{@{}l@{}}Tau Scrolls \\ Summ\\ScreenFD\end{tabular}     & \begin{tabular}[c]{@{}l@{}}It's the first day of school at Degrassi Community School, \\ and eighth-grader Ashley already has her sights set on becoming \\ the school's newest student council president. \\ Her seemingly sure win is soon threatened when her stepbrother, \\ Toby, becomes frustrated by her unchallenged status \\ and convinces his friend J.T. to run against her. \\ Meanwhile, Emma and Manny deal with eighth-grader \\ Spinner's bullying. Note: This episode marks the first... 
\end{tabular} & \begin{tabular}[c]{@{}l@{}}{[}The Kerwin House - Ashley's Room{]}\\ \\ (While getting ready for school, she's talking to \\ her friend Terri on the phone.)\\ \\ Ashley: This is gonna be the best year ever! (Working on her \\ poster for Degrassi...\end{tabular}                                                                                                                                                                                                                                                           \\ \midrule
\begin{tabular}[c]{@{}l@{}}Tau Scrolls \\ Gov. Report\end{tabular}      & \begin{tabular}[c]{@{}l@{}}Members of Congress and Administrations have periodically\\ considered reorganizing the federal government's trade and \\ development functions to advance various U.S. policy objectives.\\ The Better Utilization of Investments Leading to Development Act \\of 2018 (BUILD Act), which was signed into law on October 5, 2018\\ (P.L. 115-254), represents a potentially major overhaul of U.S.\\ development finance efforts...\end{tabular}                                                                                                                                                                                                         & \begin{tabular}[c]{@{}l@{}}Background\\  What is the U.S. International \\ Development Finance Corporation (IDFC)?\\ The IDFC is authorized by statute to be \\ a "wholly owned Government corporation ... \\ under the foreign policy guidance of the \\ Secretary of State" in the executive branch. \\ Its purpose is to "mobilize and...\end{tabular}                                                         \\ \midrule
\begin{tabular}[c]{@{}l@{}}Tau Scrolls \\ QMSUM\end{tabular}            & \begin{tabular}[c]{@{}l@{}}According to the Industrial Design, there might be only a few choices \\ for the energy source and materials from the current manufacturer, \\ so he suggested that they had better look for another manufacturer \\ for more alternatives. The Marketing put forward to design a user\\ friendly interface while the User Interface came up with the idea\\ of including the voice recognition system into the remote control...\end{tabular}                                                                                                                                                                                                            & \begin{tabular}[c]{@{}l@{}}Summarize the ideas of the individual presentations.\\ Marketing: \{vocalsound\} That went well , thank you .\\ Project Manager: That's great .\\ Industrial Designer: \{vocalsound\} 'Kay .\\ Marketing: Perfect .\\ Project Manager: Alright , let me just PowerPoint this up . \\ \{vocalsound\} \{vocalsound\} \{vocalsound\} Right so um this\end{tabular}                                                                                                                                  \\ \midrule
\begin{tabular}[c]{@{}l@{}}QASPER\\ Title\end{tabular}    & Knowledge Authoring and Question Answering with KALM                                                                                                                                                                                                                                                                                                                                                                                                                                                                                                                                                                                                                                   & \begin{tabular}[c]{@{}l@{}}Introduction: Knowledge representation and reasoning \\ (KRR) is the process of representing the \\ domain knowledge in formal languages \\ (e.g., SPARQL, Prolog) such that it can be...\end{tabular}                                                  \\ \midrule
\begin{tabular}[c]{@{}l@{}}QASPER\\ Abstract \end{tabular} & \begin{tabular}[c]{@{}l@{}}Knowledge representation and reasoning (KRR) is one of the key areas \\ in artificial intelligence (AI) field. It is intended to represent the world \\ knowledge in formal languages (e.g., Prolog, SPARQL) and then enhance \\ the expert systems to perform querying and inference tasks. Currently, \\ constructing large scale knowledge bases (KBs) with high quality is \\ prohibited by the fact that the construction\end{tabular}                                                                                                                                                                                                                 & \begin{tabular}[c]{@{}l@{}}Introduction: Knowledge representation and reasoning \\ (KRR) is the process of representing the \\ domain knowledge in formal languages \\ (e.g., SPARQL, Prolog) such that it can be... \end{tabular}                                                  \\ \midrule
\begin{tabular}[c]{@{}l@{}}MultiField\\QA\end{tabular}                                                              & What algorithm is engaged in the PLMS-PPIC method?                                                                                                                                                                                                                                                                                                                                                                                                                                                                                                                                                                                                                                     & \begin{tabular}[c]{@{}l@{}}\textbackslash{}section\{Introduction\}\textbackslash{}label\{S1\} The multiple access \\interferences (MAI) is the root of user limitation in CDMA \\ systems \textbackslash{}cite\{R1,R3\}. The parallel least mean square-partial\\ parallel interference cancelation (PLMS-PPIC) method is a \\ multiuser detector for code division multiple access (CDMA)\\ receivers which reduces the effect of\\ MAI in bit detection. In this method and\\ similar to its former version\end{tabular} \\ \midrule
2WikimQA                                                                  & Where did the director of film The Brave Bulls (Film) die?                                                                                                                                                                                                                                                                                                                                                                                                                                                                                                                                                                                                                             & \begin{tabular}[c]{@{}l@{}}Passage 1: The Brave Archer\\ The Brave Archer, also known as Kungfu Warlord,\\ is a 1977 Hong Kong film adapted from Louis\\ Cha's novel The Legend of the Condor Heroes.\\ The film was produced by the Shaw Brothers\\ Studio and directed by... \end{tabular}                                                              \\ \midrule
\begin{tabular}[c]{@{}l@{}}Passage \\ Retrieval\end{tabular}              & \begin{tabular}[c]{@{}l@{}}During World War II, navy nurses played a crucial role in \\ providing medical care and preventing further casualties. They\\ were present  during the initial  Japanese attack on Pearl Harbor,\\ as well as in Kaneohe Bay, the Philippines, Guam, and \\ aboard the Solace. The nursing profession was recognized for its \\essential contribution and was placed under the War Manpower\\ Commission. Despite shortages, ...\end{tabular}                                                                                                                                                                                                                & \begin{tabular}[c]{@{}l@{}}Paragraph 1: Thermometric titrimetry \\ Thermometric titrimetry is an\\ extraordinarily versatile technique.\\ This is differentiated from calorimetric \\ titrimetry by the fact that the heat...\end{tabular}                                                        \\ \midrule
\begin{tabular}[c]{@{}l@{}}Court\\Listener \\ (HTML)\end{tabular}           & \begin{tabular}[c]{@{}l@{}}noting that “{[}a{]}s a court of limited jurisdiction, we begin, and end, \\ with an examination of our jurisdiction”\end{tabular}                                                                                                                                                                                                                                                                                                                                                                                                                                                                                                                          &  \begin{tabular}[c]{@{}l@{}}\textless{}citances\textgreater{}[c]Sellar v Lasotav Pty Ltd: In the\\ matter of Lasotav Pty Ltd [2008] FCA 1612\\ (27 October 2008)\textless{}/citances\textgreater{} \\ 
 \textless{}/citances\textgreater{}  Home   Databases   WorldLII   Search
\end{tabular} 
\\ \midrule
\begin{tabular}[c]{@{}l@{}}Court\\Listener \\ (Plain Text)\end{tabular}     & \begin{tabular}[c]{@{}l@{}}noting that “{[}a{]}s a court of limited jurisdiction, we begin, and end, \\ with an examination of our jurisdiction”\end{tabular}                                                                                                                                                                                                                                                                                                                                                                                                                                                                                                                          & \begin{tabular}[c]{@{}l@{}}[c]Sellar v Lasotav Pty Ltd: In the\\ matter of Lasotav Pty Ltd [2008] \\ FCA 1612 (27 October 2008) \\ 
   Home   Databases   WorldLII   Search\\   Feedback  Federal Court of Australia 
\end{tabular}                                                                                                                                                                                                                                                                                                              \\ \midrule
\begin{tabular}[c]{@{}l@{}}Legal Case \\ Reports\end{tabular}             & \begin{tabular}[c]{@{}l@{}}\textless{}citphrase id="cp0.0"\textgreater{}cited from="{[}2006{]} FCA 1222"\&\\gt;corporations law\textless{}/citphrase\textgreater\\ \textless{}citphrase id="cp0.1"\textgreater{}cited from="{[}2006{]} FCA 1222"\&gt;\\ funds held pursuant  to terminated deed of company arrangement \\ are held for the benefit of deed creditors...\end{tabular}                                                                                 & \begin{tabular}[c]{@{}l@{}}On 14 November 2008, Ms Swee Yen Tay\\ instituted a proceeding \\ in this Court against the Migration Review\\ Tribunal ("the Tribunal") \\ and the Minister for Immigration\\ and Citizenship.\textless{}/sentence\textgreater\\ \end{tabular}           \\ \midrule
\begin{tabular}[c]{@{}l@{}}Stack\\ Overflow\end{tabular}                  & \begin{tabular}[c]{@{}l@{}}Multithreading Design Best Practice | Consider this problem: I have \\ a program which should fetch (let's say) 100 records from a database,\\ and then for each one it should get updated information from a web\\ service. There are two ways to introduce parallelism in this scenario:\end{tabular}                                                                                                                                                                                                         & \begin{tabular}[c]{@{}l@{}}You could use an Observer pattern. A simple functional way \\to accomplish this: ``` \textless{}php  Plugin system  listeners = array(); \\ Create an entry... \end{tabular} \\ \bottomrule                                                                                                                                                                                                                                                                                                        
\end{tabular}
\caption{LoCoV1 Examples for each Dataset}
\label{tab:loco_examples}
\end{table}
\clearpage

\subsection{M2-BERT Pretraining, Fine-Tuning,  and Evaluation Details}
\label{sec:m2_training_finetuning_eval}

For pretraining the M2-BERT encoders, we use the C4, Wikipedia, and Bookcorpus datasets for training examples.
For our dataset split, we sample each dataset equally (e.g. 33\% each). 
For our example length ratio, we selected 0.3 variable length examples (e.g. short examples) and 0.7 maximum concatenated examples (e.g. long examples).
We utilize the masked-language modeling (MLM) pretraining objective with an MLM probability of 0.3 to prepare the encoders for downstream language modeling.
For training evaluation, we use the C4 validation set with an MLM probability of 0.15.
For our scheduler, we use linear decay with warmup, where warmup is 0.06 of the total training duration.
For our optimizer, we use a learning rate of $5.0e-4$ with an epsilon of $1e-06$, betas of $0.9$ and $0.98$, a weight decay of $1e-5$.

For fine-tuning the M2-BERT encoders, we use the Sentence Transformers library \cite{Reimers2019SentenceBERTSE}. 
For all M2-BERT configurations, we use a learning rate of $5e-6$, a true batch size of 32, 1 epoch of fine-tuning, a maximum gradient norm of 1.0, and a ratio of 32 negative passages per query-positive passage pair.
When using orthogonal projection loss (OPL) for fine-tuning, we use cosine similarity distance for calculating loss.
When using prototype loss (PL), we first fine-tune the M2-BERT-32k model with the fine-tuned M2-BERT-128 model as the teacher model.
To improve downstream retrieval accuracy, we then have a second-phase of fine-tuning in which we fine-tune with MNRL with a batch size of 2.

For evaluation, we use the BEIR library \cite{thakur2021beir} to calculate retrieval accuracy on both the LoCoV1 and BEIR benchmarks.
For accuracy, we use normalized discounted cumulative gain at 10 (nDCG@10) \cite{wang2013theoretical}.
\clearpage

\subsection{M2-BERT on BEIR - Expanded Results}

\begin{table}[h!]
\small
\centering
\begin{tabular}{lcc}
\toprule
\textbf{Model}           & \begin{tabular}[c]{@{}c@{}}sentence-transformers/\\ msmarco-bert-base-dot-v5\end{tabular} & M2-BERT \\ \midrule
\textbf{Max Seq. Length} & 512                                                                                       & 128     \\ \midrule
\textbf{Param. Count}    & 110M                                                                                      & 80M     \\ \midrule
MSMARCO                  & 65.0                                                                                      & 59.8    \\
TREC COVID               & 35.8                                                                                      & 43.3    \\
NFCorpus                 & 23.1                                                                                      & 24.6    \\
NQ                       & 34.5                                                                                      & 30.6    \\
HotPot QA                & 44.7                                                                                      & 39.8    \\
FIQA                     & 22.0                                                                                      & 22.7    \\
Arguana                  & 42.1                                                                                      & 42.0    \\
Webis Touche 2020        & 11.0                                                                                      & 19.1    \\
Quora                    & 84.6                                                                                      & 84.2    \\
DBpedia Entity           & 30.1                                                                                      & 28.5    \\
SciDocs                  & 14.5                                                                                      & 10.9    \\
Climate Fever            & 55.4                                                                                      & 57.8    \\
SciFact                  & 56.9                                                                                      & 39.9 \\ \toprule
\textbf{BEIR Score Average} & \textbf{64.5} & 63.4 \\
\bottomrule   
\end{tabular}
\label{tab:m2_vs_sbert_expanded}
\caption{Expanded Results for M2-BERT Retrieval Encoder vs. SentenceBERT on BEIR.}
\end{table}
\clearpage

\subsection{M2-BERT on LoCoV1 - Expanded Results}

\begin{table}[h!]
\tiny
\centering
\setlength{\tabcolsep}{1.0pt}
\begin{tabular}{cccccccccccccccc}
\toprule
\multicolumn{1}{l}{}                                                            & \multicolumn{1}{l}{}                                          & \multicolumn{1}{l}{}                                                & \multicolumn{12}{c}{\textbf{LoCoV1 Datasets}}                                                                                                                                                                                                                                                                                                                                                                                                                                                                                                                                                                                                                                                                                                                                                                                                                                                     & \multicolumn{1}{l}{}                                          \\ \midrule
\textbf{Model}                                                                  & \textbf{\begin{tabular}[c]{@{}c@{}} Param. \\ Count \end{tabular}} & \textbf{\begin{tabular}[c]{@{}c@{}}MaxSeq\\ Length\end{tabular}} & \textbf{\begin{tabular}[c]{@{}c@{}}Summ\\ Screen\\FD\end{tabular}} & \textbf{\begin{tabular}[c]{@{}c@{}}Gov. \\Report\end{tabular}} & \textbf{\begin{tabular}[c]{@{}c@{}}QM\\SUM\end{tabular}} & \textbf{\begin{tabular}[c]{@{}c@{}}QASPER\\ Title\end{tabular}} & \textbf{\begin{tabular}[c]{@{}c@{}}QASPER\\ Abstract\end{tabular}} & \textbf{\begin{tabular}[c]{@{}c@{}}Multi\\Field\\ QA\end{tabular}} & \textbf{\begin{tabular}[c]{@{}c@{}}2Wiki \\ MQA\end{tabular}} & \textbf{\begin{tabular}[c]{@{}c@{}}Passage \\ Retrieval\end{tabular}} & \textbf{\begin{tabular}[c]{@{}c@{}}Court\\Listener \\ (HTML)\end{tabular}} & \textbf{\begin{tabular}[c]{@{}c@{}}Court\\Listener \\ (Text)\end{tabular}} & \textbf{\begin{tabular}[c]{@{}c@{}}Legal\\ Case\\ Reports\end{tabular}} & \textbf{\begin{tabular}[c]{@{}c@{}}S.O.\end{tabular}} & \textbf{\begin{tabular}[c]{@{}c@{}}LoCo \\ Avg.\end{tabular}} \\ \midrule
\begin{tabular}[c]{@{}c@{}}BGE-Large\\ Zero-Shot\end{tabular}                   & 335M                                                          & 512                                                                 & 65.8                                                                              & 92.6                                                                          & 46.0                                                                    & 87.1                                                                               & 94.5                                                                                  & 89.3                                                             & 69.4              & 20.1                                                                  & 10                                                                       & 10.1                                                                           & 18.9                                                                  & 74.1                                                              & 56.5                                                          \\ \midrule
\begin{tabular}[c]{@{}c@{}}BGE-Large\\ Fine-tuned\end{tabular}                  & 335M                                                          & 512                                                                 & 84.8                                                                              & 96.0                                                                          & 67.9                                                                    & 93.5                                                                               & 97.8                                                                                  & 92.1                                                             & 71.1              & 22.5                                                                  & 22.0                                                                     & 22.8                                                                           & 42.6                                                                  & 76.5                                                              & 64.8                                                          \\ \midrule
\begin{tabular}[c]{@{}c@{}}BGE-Large \\ Fine-tuned \\ w. Chunks\end{tabular}  & 335M                                                          & 512                                                                 & 80.7                                                                              & 95.5                                                                          & 58.1                                                                    & 89.3                                                                               & 96.6                                                                                  & 88.4                                                             & 66.4              & 20.3                                                                  & 21.0                                                                     & 21.8                                                                           & 39.9                                                                  & 76.7                                                              & 61.6                                                          \\ \midrule
E5-Mistral                & 7.11B                                                         & 4096                                                                & 95.9                                                                              & 98.3                                                                          & 46.8                                                                    & \textbf{98.4}                                                                               & \textbf{99.8}                                                                                  & 93.5                                                             & 88.3              & 35.3                                                                  & 33.9                                                                     & 34.6                                                                           & 49.5                                                                  & 82.7                                                              & 71.4                                                          \\ \midrule
\begin{tabular}[c]{@{}c@{}}E5-Mistral \\ w. Chunks\end{tabular} & 7.11B                                                         & 4096                                                                & 95.6                                                                              & 98.4                                                                          & 47.6                                                                    & 96.8                                                                               & 99.7                                                                                  & 90.5                                                             & 84.8              & 32.9                                                                  & 32.8                                                                     & 32.7                                                                           & 49.2                                                                  & 83.1                                                              & 70.3                                                          \\ \midrule
Jina Embeds.                                                                    & 137M                                                          & 8192                                                                & 93.3                                                                              & 98.6                                                                          & 40.5                                                                    & 95.1                                                                               & 99.4                                                                                  & 86.4                                                             & 81.6              & 60.7                                                                  & 27.0                                                                     & 26.1                                                                           & 30.7                                                                  & 69.0                                                              & 67.2                                                          \\ \midrule
\begin{tabular}[c]{@{}c@{}}Jina Embeds.\\ w. Chunks\end{tabular}                & 137M                                                          & 8192                                                                & 6.1                                                                               & 25.2                                                                          & 4.2                                                                     & 32.5                                                                               & 54.3                                                                                  & 43.8                                                             & 21.6              & 10.4                                                                  & 0.9                                                                      & 0.5                                                                            & 1.8                                                                   & 28.9                                                              & 19.2                                                          \\ \midrule
OpenAI Ada                                                                      & N/A                                                           & 8192                                                                & 86.2                                                                              & 97.1                                                                          & 48.8                                                                    & 93.8                                                                               & 98.9                                                                                  & 90.1                                                             & 78.9              & 31.2                                                                  & 16.3                                                                     & 16.8                                                                           & 28.2                                                                  & 72.3                                                              & 63.2                                                          \\ \midrule
\begin{tabular}[c]{@{}c@{}}OpenAI Ada\\ w. Chunks\end{tabular}                  & N/A                                                           & 8192                                                                & 86.2                                                                              & 97.1                                                                          & 49.0                                                                    & 93.8                                                                               & 98.9                                                                                  & 90.1                                                             & 78.9              & 31.2                                                                  & 16.3                                                                     & 16.8                                                                           & 30.7                                                                  & 72.3                                                              & 63.4  
\\ \midrule
\begin{tabular}[c]{@{}c@{}}ColBERTv2\end{tabular}                  & 110M                                                           & 512                                                                & 66.5	& 88.0 &	36.2 &	85.5 &	94.5 & 85.0	& 71.7 &	21.5 & 14.7 & 17.6 & 17.2 & 44.5  & 53.6                                                      \\ \midrule
Voyage-001                                                                      & N/A                                                           & 4096                                                                & 76.7                                                                              & 92.4                                                                          & 52.9                                                                    & 88.4                                                                               & 91.7                                                                                  & 88.7                                                             & 57.0              & 17.7                                                                  & 13.0                                                                     & 12.8                                                                           & 14.0                                                                  & 74.9                                                              & 56.7                                                          \\ \midrule
\begin{tabular}[c]{@{}c@{}}Cohere \\ Embed-Eng.\\ v3.0\end{tabular}             & N/A                                                           & 512                                                                 & 75.3                                                                              & 92.2                                                                          & 38.1                                                                    & 89.8                                                                               & 93.1                                                                                  & 88.9                                                             & 68.2              & 22.1                                                                  & 13.3                                                                     & 14.3                                                                           & 24.3                                                                  & 75.3                                                              & 57.9                                                          \\ \midrule
M2-BERT                                                                         & 80M                                                           & 128                                                                 & 64.4                                                                              & 85.3                                                                          & 61.2                                                                    & 77.2                                                                               & 83.2                                                                                  & 91.4                                                             & 76.3              & 39.7                                                                  & 82.5                                                                     & 84.8                                                                           & 26.9                                                                  & 69.0                                                              & 70.1                                                          \\ \midrule
M2-BERT                                                                         & 80M                                                           & 2048                                                                & 78.5                                                                              & 94.4                                                                          & 69.2                                                                    & 88.1                                                                               & 96.6                                                                                  & 93.4                                                             & 83.0              & 69.9                                                                  & 92.0                                                                     & 92.0                                                                           & 51.5                                                                  & 79.6                                                              & 82.3                                                          \\ \midrule
M2-BERT                                                                         & 80M                                                           & 8192                                                                & 84.6                                                                              & 96.5                                                                          & 69.6                                                                    & 93.1                                                                               & 98.9                                                                                  & 97.1                                                             & 87.7              & 82.0                                                                  & 94.9                                                                     & 95.4                                                                           & 58.6                                                                  & 84.8                                                              & 86.9                                                          \\ \midrule
M2-BERT                                                                         & 80M                                                           & 32768                                                         & \textbf{98.0}  & \textbf{98.7} & \textbf{70.4} & 97.9 & 98.3 & \textbf{98.5} & \textbf{92.1} & \textbf{89.3} & \textbf{96.8} & \textbf{97.0} & \textbf{67.1} & \textbf{88.9} & \textbf{91.1} \\ \bottomrule                                                             
\end{tabular}
\caption{M2-BERT and Baseline Model Performances on LoCoV1 benchmark - Complete Results.}
\label{tab:m2_on_loco_complete_results}
\end{table}
\clearpage

\subsection{BEIR Dataset Examples}

\begin{table}[h!]
\centering
\tiny
\setlength{\tabcolsep}{2.0pt}
\begin{tabular}{lll}
\toprule
\textbf{BEIR Dataset} & \textbf{Query Example} & \textbf{Document Example} \\ \midrule
SciFact & 1/2000 in UK have abnormal PrP positivity. & \begin{tabular}{@{}l@{}}OBJECTIVES To carry out a further survey of archived \\ appendix samples to understand better\\  the differences between existing estimates of the prevalence \\of subclinical infection with prions after \\ the bovine spongiform encephalopathy epizootic \\and to see whether a broader birth cohort was \\ affected, and to understand better the \\implications for the management of blood and blood products \\ and for the handling of surgical instruments.\\ DESIGN Irreversibly unlinked and anonymised large \\ scale survey of archived appendix samples.\\ SETTING Archived appendix samples from the pathology \\ departments of 41 UK hospitals participating\\ in the earlier survey, and additional hospitals \\ in regions with lower levels of participation in that\\ survey. SAMPLE 32,441 archived appendix \\ samples fixed in formalin and embedded in paraffin and\\ tested for the presence of abnormal prion protein \\ (PrP). RESULTS Of the 32,441 appendix samples 16 were\\ positive for abnormal PrP, indicating an overall \\ prevalence of 493 per million population (95\% confidence\\ interval 282 to 801 per million). The prevalence \\ in those born in 1941-60 (733 per million, 269 to\\ 1596 per million) did not differ significantly \\ from those born between 1961 and 1985 (412 per million,\\ 198 to 758 per million) and was similar in \\ both sexes and across the three broad geographical areas\\ sampled. Genetic testing of the positive \\ specimens for the genotype at PRNP codon 129 revealed a\\ high proportion that were valine homozygous \\ compared with the frequency in the normal population,\\ and in stark contrast with confirmed \\ clinical cases of vCJD, all of which were methionine\\ homozygous at PRNP codon 129. CONCLUSIONS \\ This study corroborates previous studies and suggests a\\ high prevalence of infection with \\ abnormal PrP, indicating vCJD carrier status in the population.\end{tabular} \\ \midrule
Quora & \begin{tabular}{@{}l@{}}How do Russian politics and geostrategy \\ affect Australia and New Zealand?\end{tabular} & How does Russian politics affect Australia and New Zealand? \\ \midrule
NQ & where does junior want to go to find hope & \begin{tabular}{@{}l@{}}Throughout the novel, Junior shares his dreams with the readers.\\ In the first chapter, he dreams of becoming a cartoon artist in order to\\ get rich and escape the cycles of poverty and abuse on the reservation\\ The idea that hope exists off the rez is\\ echoed in later chapters, where Junior finds himself caught between\\  home on the reservation and pursuing his dreams in the outside world.\\ Junior asks his parents, "Who has the most\\ hope?" to which they answer "White people".[h] The rez\\ is characterized by lack of opportunity and poor education,\\ the solution to which appears to lie in the Western world. \end{tabular} \\ \midrule
MSMARCO & cost of interior concrete flooring & \begin{tabular}{@{}l@{}}For a 4 inch concrete floor, 1 yard of concrete will cover 80 square feet.\\ The cost would be very close either way for a 4 inch concrete floor.\\ If the floor is thicker than 4 inches, then the surface hardener\\ is less money to use.\end{tabular} \\ \midrule
TREC-COVID & \begin{tabular}{@{}l@{}} how does the coronavirus respond \\to changes in the weather\end{tabular} & \begin{tabular}{@{}l@{}}Abstract In this study, we aimed at analyzing the associations between transmission\\ of and deaths caused by SARS-CoV-2 and meteorological variables, such as average \\temperature, minimum temperature, maximum temperature, and precipitation.\\ Two outcome measures were considered, with the\\ first aiming to study SARS-CoV-2 infections \\and the second aiming to study COVID-19 \\mortality. Daily data as well as data on SARS-CoV-2 infections and\\ COVID-19 mortality obtained between December 1, 2019 and\\ March 28, 2020 were collected from weather \\stations around the world. The country's\\ population density and time of exposure to the disease were\\ used as control variables. Finally, a month dummy\\ variable was added. Daily data by country\\ were analyzed using the panel data model.\\ An increase in the average daily temperature by one degree \\Fahrenheit reduced the number of cases by approximately \\6.4 cases/day. There was a negative correlation\\ between the average temperature per\\ country and the number of cases of SARS-CoV-2 infections...\end{tabular} \\
\bottomrule
\end{tabular}
\caption{BEIR Benchmark Examples}
\label{tab:beir_examples}
\end{table}
\clearpage

\subsection{Needle-in-the-Haystack Synthetic Task - Expanded Results}

\begin{table}[H]
\tiny
\centering
\begin{tabular}{cccccc}
\toprule
\textbf{Model}                                                                        & BGE-Large Zero-shot & E5-Mistral & Jina Embeds. & OpenAI Ada Embeds. & M2-BERT \\ \midrule
\textbf{Max. Seq. Length}                                                             & 512                                                           & 4096                                                             & 8192                                                   & 8192                                                          & 32768                       \\ \midrule
\textbf{Param. Count}                                                                 & 335M                                                          & 7.11B                                                            & 137M                                                   & N/A                                                           & 80M                         \\ \midrule
\textbf{\begin{tabular}[c]{@{}c@{}}Answer Position\\ in Concat. Passage\end{tabular}} &                                                               &                                                                  &                                                        &                                                               &                             \\ \toprule
0                                                                                     & 76.7                                                          & 68.4                                                             & 4.7                                                    & 77.0                                                          & 82.0                        \\ \midrule
1                                                                                     & 66.5                                                          & 60.2                                                             & 3.1                                                    & 60.1                                                          & 79.9                        \\ \midrule
2                                                                                     & 62.3                                                          & 42.1                                                             & 2.7                                                    & 45.0                                                          & 77.0                        \\ \midrule
3                                                                                     & 58.1                                                          & 23.9                                                             & 2.2                                                    & 30.9                                                          & 78.4                        \\ \midrule
4                                                                                     & 55.5                                                          & 10.4                                                             & 2.1                                                    & 22.0                                                          & 76.3                        \\ \midrule
5                                                                                     & 50.9                                                          & 8.1                                                              & 1.4                                                    & 15.2                                                          & 75.8                        \\ \midrule
6                                                                                     & 49.5                                                          & 6.7                                                              & 1.3                                                    & 11.1                                                          & 75.6                        \\ \midrule
7                                                                                     & 45.7                                                          & 5.4                                                              & 1.1                                                    & 8.3                                                           & 74.9                        \\ \midrule
8                                                                                     & 42.8                                                          & 4.8                                                              & 1.2                                                    & 5.5                                                           & 73.2                        \\ \midrule
9                                                                                     & 37.3                                                          & 4.5                                                              & 1.1                                                    & 4.0                                                           & 72.1                        \\ \midrule
10                                                                                    & 35.6                                                          & 3.9                                                              & 0.9                                                    & 3.2                                                           & 71.5                        \\ \midrule
11                                                                                    & 30.0                                                          & 3.2                                                              & 0.9                                                    & 2.9                                                           & 70.4                        \\ \midrule
12                                                                                    & 27.1                                                          & 3.1                                                              & 0.7                                                    & 2.1                                                           & 68.9                        \\ \midrule
13                                                                                    & 23.0                                                          & 2.5                                                              & 0.8                                                    & 1.6                                                           & 68.5                        \\ \midrule
14                                                                                    & 19.4                                                          & 1.8                                                              & 0.9                                                    & 1.6                                                           & 66.2                        \\ \midrule
15                                                                                    & 16.1                                                          & 2.7                                                              & 0.7                                                    & 1.4                                                           & 64.0                        \\ \midrule
16                                                                                    & 13.5                                                          & 2.2                                                              & 0.5                                                    & 0.9                                                           & 62.4                        \\ \midrule
17                                                                                    & 12.1                                                          & 1.9                                                              & 0.4                                                    & 1.1                                                           & 55.4                        \\ \midrule
18                                                                                    & 9.8                                                           & 1.4                                                              & 0.2                                                    & 1.0                                                           & 45.2                        \\ \midrule
19                                                                                    & 7.2                                                           & 0.9                                                              & 0.1                                                    & 1.1                                                           & 38.3                        \\ \midrule
20                                                                                    & 5.5                                                           & 1.1                                                              & 0.2                                                    & 1.0                                                           & 30.1                        \\ \midrule
21                                                                                    & 3.8                                                           & 1.4                                                              & 0.1                                                    & 1.0                                                           & 28.2                        \\ \midrule
22                                                                                    & 2.9                                                           & 0.8                                                              & 0.1                                                    & 1.0                                                           & 26.7                        \\ \midrule
23                                                                                    & 2.8                                                           & 1.2                                                              & 0.1                                                    & 0.7                                                           & 25.3                        \\ \midrule
24                                                                                    & 2.2                                                           & 1.3                                                              & 0.4                                                    & 0.8                                                           & 21.3                        \\ \midrule
25                                                                                    & 1.7                                                           & 1.0                                                              & 0.1                                                    & 1.0                                                           & 20.5                        \\ \midrule
26                                                                                    & 1.3                                                           & 1.0                                                              & 0.2                                                    & 0.8                                                           & 19.2                        \\ \midrule
27                                                                                    & 1.1                                                           & 1.3                                                              & 0.2                                                    & 0.7                                                           & 17.9                        \\ \midrule
28                                                                                    & 1.1                                                           & 0.8                                                              & 0.1                                                    & 0.8                                                           & 16.9                        \\ \midrule
29                                                                                    & 1.0                                                           & 1.3                                                              & 0.2                                                    & 1.0                                                           & 15.3                        \\ \midrule
30                                                                                    & 1.0                                                           & 1.2                                                              & 0.1                                                    & 1.2                                                           & 12.3                        \\ \midrule
31                                                                                    & 0.7                                                           & 0.8                                                              & 0.1                                                    & 1.1                                                           & 8.3                         \\ \midrule
32                                                                                    & 0.8                                                           & 0.4                                                              & 0.3                                                    & 1.3                                                           & 6.1                         \\ \midrule
33                                                                                    & 1.0                                                           & 0.5                                                              & 0.3                                                    & 1.8                                                           & 5.5                         \\ \midrule
34                                                                                    & 0.1                                                           & 0.6                                                              & 0.1                                                    & 2.1                                                           & 3.2                         \\ \midrule
35                                                                                    & 0.1                                                           & 0.3                                                              & 0.2                                                    & 1.7                                                           & 1.9                         \\ \midrule
36                                                                                    & 0.2                                                           & 0.5                                                              & 0.3                                                    & 2.1                                                           & 1.4                         \\ \midrule
37                                                                                    & 0.2                                                           & 0.4                                                              & 0.1                                                    & 3.1                                                           & 1.0                         \\ \midrule
38                                                                                    & 0.2                                                           & 0.4                                                              & 0.1                                                    & 3.5                                                           & 1.1                         \\ \midrule
39                                                                                    & 0.1                                                           & 0.3                                                              & 0.1                                                    & 3.8                                                           & 0.8                         \\ \toprule
\textbf{Synth. Task Avg.}                                                                      & 19.2                                                          & 6.9                                                              & 0.8                                                    & 8.2                                                           & \textbf{41.0}     \\ \bottomrule                  
\end{tabular}
\caption{M2-BERT and Baseline Performances on Needle-in-the-Haystack Synthetic Task - Complete Results.}
\label{tab:m2_on_synth_task_complete_results}
\end{table}

\clearpage

\subsection{MTEB Benchmark Results}
\label{sec:mteb_benchmark_complete_results}

\begin{table}[H]
\centering 
\small
\begin{tabular}{ccc}
\toprule
\textbf{Model}                         & SentenceBERT & M2-BERT \\ \midrule
\textbf{Max. Seq. Length}              & 512          & 128     \\ \midrule
\textbf{Param. Count}                  & 110M         & 80M     \\ \midrule
AmazonCounterfactualClassification     & 66.0         & 66.7    \\ \midrule
AmazonPolarityClassification           & 63.8         & 73.4    \\ \midrule
AmazonReviewsClassification            & 32.5         & 37.5    \\ \midrule
Banking77Classification                & 81.2         & 78.2    \\ \midrule
EmotionClassification                  & 44.3         & 42.8    \\ \midrule
ImdbClassification                     & 59.7         & 60.4    \\ \midrule
MassiveIntentClassification            & 68.4         & 63.5    \\ \midrule
MassiveScenarioClassification          & 73.1         & 71.6    \\ \midrule
MTOPDomainClassification               & 91.4         & 85.1    \\ \midrule
MTOPIntentClassification               & 71.9         & 59.2    \\ \midrule
ToxicConversationsClassification       & 66.9         & 65.0    \\ \midrule
TweetSentimentExtractionClassification & 54.8         & 57.6    \\ \toprule
\textbf{Average Accuracy}  & \textbf{64.5}         & 63.4 \\ \bottomrule  
\end{tabular}
\caption{M2-BERT-128 and SentenceBERT Performance on MTEB Classification - Complete Results.}
\label{tab:mteb_classification}
\end{table}

\begin{table}[H]
\centering 
\small
\begin{tabular}{ccc}
\toprule
\textbf{Model}            & SentenceBERT & M2-BERT \\ \midrule
\textbf{Max. Seq. Length} & 512          & 128     \\ \midrule
\textbf{Param. Count}     & 110M         & 80M     \\ \midrule
ArXiv Clustering P2P      & 37.3         & 31.9    \\
ArXiv Clustering S2S      & 25.9         & 25.7    \\
BiorxivClusteringP2P      & 31.6         & 27.53   \\
BiorxivClusteringS2S      & 25.2         & 23.4    \\
MedrxivClusteringP2P      & 28.8         & 27.6    \\
MedrxivClusteringS2S      & 25.0           & 26.4    \\
RedditClustering          & 42.5         & 47.6    \\
RedditClusteringP2P       & 53.3         & 49.9    \\ \midrule
\textbf{Average V. Measure} & \textbf{33.7}         & 32.5  \\ \bottomrule
\end{tabular}
\caption{M2-BERT-128 and SentenceBERT Performance on MTEB Clustering - Complete Results.}
\label{tab:mteb_clustering}
\end{table}

\begin{table}[H]
\centering 
\small
\begin{tabular}{ccc}
\toprule
\textbf{Model}            & SentenceBERT  & M2-BERT \\ \midrule
\textbf{Max Seq. Length}  & 512           & 128     \\ \midrule
\textbf{Param. Count}     & 110M          & 80M     \\ \midrule
SprintDuplicateQuestions  & 99.7          & 99.8    \\
TwitterSemEval2015        & 84            & 82.9    \\
TwitterURLCorpus          & 87.9          & 88.1    \\ \midrule
\textbf{Average Cosine Similarity} & \textbf{90.5} & 90.3   \\ \bottomrule
\end{tabular}
\caption{M2-BERT-128 and SentenceBERT Performance on MTEB Pair Classification - Complete Results.}
\label{tab:mteb_pair_classification}
\end{table}

\begin{table}[H]
\centering 
\small
\begin{tabular}{ccc}
\toprule
\textbf{Model}            & SentenceBERT & M2-BERT       \\ \midrule
\textbf{Max Seq. Length}  & 512          & 128           \\ \midrule
\textbf{Param. Count}     & 110M         & 80M           \\ \midrule
AskUbuntuDupQuestions     & 56.4         & 57.8          \\
MindSmallReranking        & 29.6         & 30.8          \\
SciDocsRR                 & 70.7         & 71.6          \\
StackOverflowDupQuestions & 46.8         & 45.0            \\ \midrule
\textbf{Average MAP}      & 50.9         & \textbf{51.3} \\ \bottomrule
\end{tabular}
\caption{M2-BERT-128 and SentenceBERT Performance on MTEB Reranking - Complete Results.}
\label{tab:mteb_reranking}
\end{table}

\begin{table}[H]
\centering 
\small
\begin{tabular}{ccc}
\toprule
\textbf{Model}                                                                                   & SentenceBERT & M2-BERT       \\ \midrule
\textbf{Max Seq. Length}                                                                         & 512          & 128           \\ \midrule
\textbf{Param. Count}                                                                            & 110M         & 80M           \\ \midrule
BIOSSES                                                                                          & 84.9         & 84.5          \\ 
SICK-R                                                                                           & 75.7         & 82.4          \\
STS12                                                                                            & 69           & 79.7          \\
STS13                                                                                            & 75.3         & 75.9          \\
STS14                                                                                            & 74.1         & 78.3          \\
STS15                                                                                            & 81.3         & 80.1          \\
STS16                                                                                            & 76.7         & 78.4          \\
STS17                                                                                            & 83.7         & 82.8          \\
STS22                                                                                            & 63.4         & 64.7          \\
STSBenchmark                                                                                     & 76.8         & 81.3          \\ \midrule
\textbf{\begin{tabular}[c]{@{}c@{}}Average Pearson Corr.\\ for Cosine Similarities\end{tabular}} & 76.1         & \textbf{78.8} \\ \bottomrule
\end{tabular}
\caption{M2-BERT-128 and SentenceBERT Performance on MTEB STS - Complete Results.}
\label{tab:mteb_sts}
\end{table}

\clearpage

\subsection{M2-BERT Pretraining Strategies}

\begin{figure}[ht]
   \centering
   \includegraphics[width=0.7\linewidth]{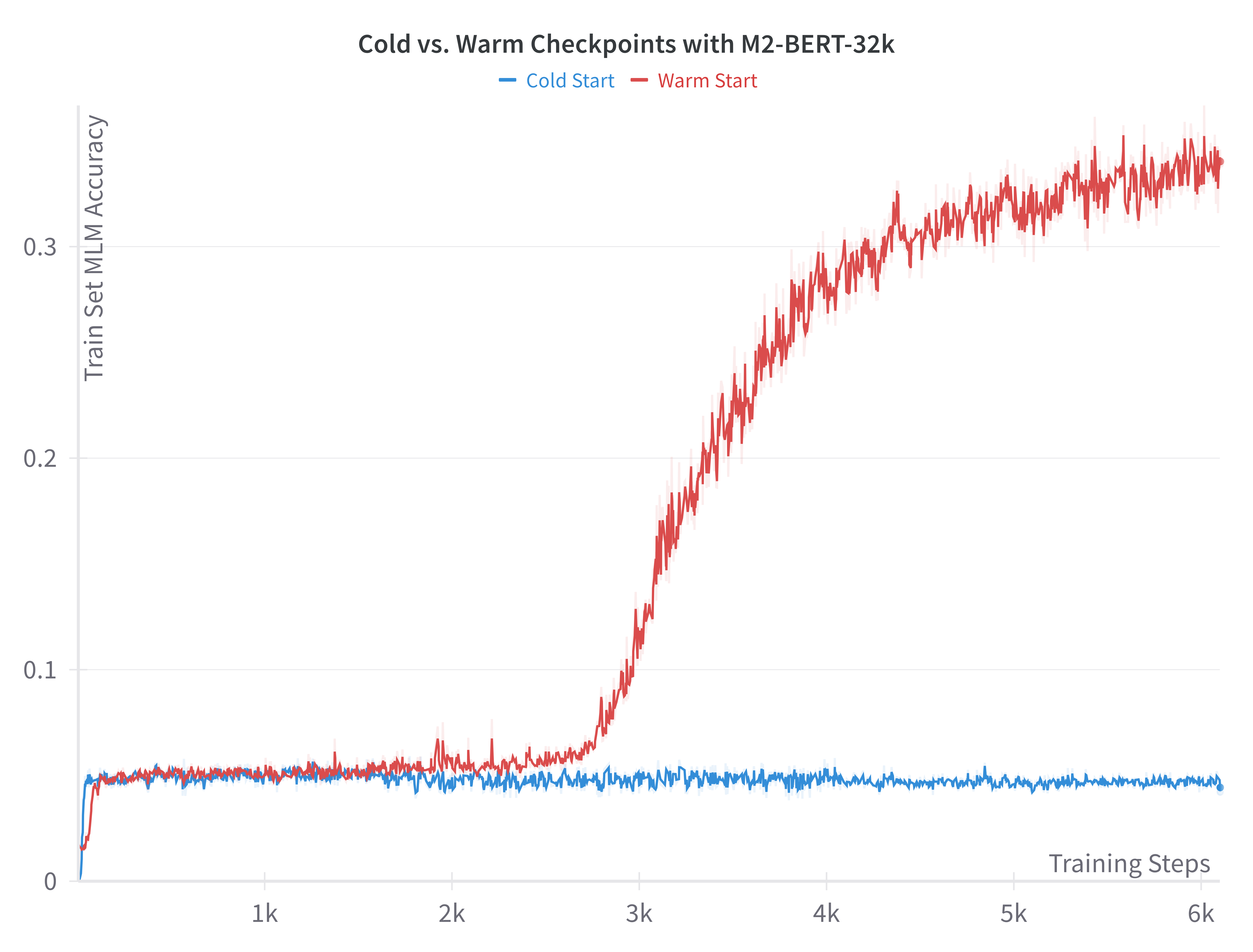}
   \caption{Cold vs. Warm Start for M2-BERT-32k Pretraining Checkpoints.}
   \label{fig:warm_vs_cold_checkpoints_graph}
\end{figure}
\clearpage

\subsection{Baseline Model Selection}\label{sec:model_selection_hyperlinks}

\begin{itemize}
    \item BGE-Large-en-v1.5: \hyperlink{https://huggingface.co/BAAI/bge-large-en-v1.5}{https://huggingface.co/BAAI/bge-large-en-v1.5}
    \item E5-Mistral: \hyperlink{https://huggingface.co/intfloat/e5-mistral-7b-instruct}{https://huggingface.co/intfloat/e5-mistral-7b-instruct}
    \item Jina Embeddings: \hyperlink{https://huggingface.co/jinaai/jina-embeddings-v2-base-en}{https://huggingface.co/jinaai/jina-embeddings-v2-base-en}
    \item OpenAI Ada Embeddings: \hyperlink{https://platform.openai.com/docs/guides/embeddings}{https://platform.openai.com/docs/guides/embeddings}
    \item VoyageAI Voyage-001 Embeddings: \hyperlink{https://docs.voyageai.com/embeddings/}{https://docs.voyageai.com/embeddings/}
    \item Cohere Embed-English v3.0: \hyperlink{https://cohere.com/models/embed}{https://cohere.com/models/embed}
    \item Okapi BM25: \hyperlink{https://www.elastic.co/}{https://www.elastic.co/}
\end{itemize}
\clearpage

\subsection{M2-BERT Efficiency Experiments}\label{sec:efficiency_details}

For all our efficiency experiments, we run each of the models on a single A100 GPU with 80GB of memory, running CUDA 11.7, Python 3.10, and PyTorch 1.13.1 \cite{paszke2019pytorch}.
We pre-tokenize all input sequences before measuring the time it takes to tokenize the entirety of the sequence, which can involve embedding separate chunks of the sequence if the model's maximum sequence length is less than the total sequence length.
\clearpage

\subsection{LoCoV0 Performance}

\begin{table}[H]
\tiny
\centering
\begin{tabular}{ccccccccc}
\toprule
                                                                    &                                                                  &                                                                     & \multicolumn{5}{c}{\textbf{LoCoV0 Dataset}}                                                                                                                                                                                                  &                                                                      \\ \midrule
\textbf{Model}                                                      & \textbf{\begin{tabular}[c]{@{}c@{}}Param.\\ Count.\end{tabular}} & \textbf{\begin{tabular}[c]{@{}c@{}}Max. Seq.\\ Length\end{tabular}} & \begin{tabular}[c]{@{}c@{}}Summ\\ ScreenFD\end{tabular} & \begin{tabular}[c]{@{}c@{}}Gov.\\ Report\end{tabular} & QMSUM & \begin{tabular}[c]{@{}c@{}}QASPER\\ Title\end{tabular} & \begin{tabular}[c]{@{}c@{}}QASPER\\ Abstract\end{tabular} & \textbf{\begin{tabular}[c]{@{}c@{}}Average\\ Score\end{tabular}} \\ \midrule
E5-Mistral                                                          & 7.11B                                                            & 4096                                                                & 95.9                                                    & 98.3                                                  & 46.8  & 98.4                                                   & 99.8                                                      & 87.8                                                                 \\ \midrule
\begin{tabular}[c]{@{}c@{}}BGE-Large\\ Fine-tuned\end{tabular}      & 335M                                                             & 512                                                                 & 70.8                                                    & 93.5                                                  & 66.0  & 96.3                                                   & 98.4                                                      & 85.0                                                                 \\ \midrule

Jina Embeds. & 137M & 8192 & 93.3 & 98.6 & 40.5 & 95.1 & 99.4 & 85.4

\\ \midrule
OpenAI Ada                                                          & N/A                                                              & 8192                                                                & 86.2                                                    & 97.1                                                  & 48.8  & 93.8                                                   & 98.9                                                      & 85.0                                                                 \\ \midrule
\begin{tabular}[c]{@{}c@{}}Cohere Embed\\ English v3.0\end{tabular} & N/A                                                              & 512                                                                 & 75.3                                                    & 92.2                                                  & 38.1  & 89.8                                                   & 93.1                                                      & 77.7                                                                \\ \midrule
\begin{tabular}[c]{@{}c@{}}Voyage\\ voyage-01\end{tabular}          & N/A                                                              & 4096                                                                & 76.7                                                    & 92.4                                                  & 52.9  & 88.4                                                   & 91.7                                                      & 80.4                                                               \\ \midrule
M2-BERT-2k                                                          & 80M                                                              & 2048                                                                & 81.8                                                    & 94.7                                                  & 58.5  & 87.3                                                   & 95.5                                                      & 83.6                                                                 \\ \midrule
M2-BERT-8k                                                          & 80M                                                              & 8192                                                                & 94.7                                                    & 96.5                                                  & 64.1  & 86.8                                                   & 97.5                                                      & 85.9                                                                 \\ \midrule
M2-BERT-32k                                                         & 80M                                                              & 32768                                                               & 98.6                                                    & 98.5                                                  & 69.5  & 97.4                                                   & 98.7                                                      & 92.5                   \\ \bottomrule                                             
\end{tabular}
\caption{M2-BERT and Baseline Model Performances on LoCoV0}
\label{tab:loco_v0_results}
\end{table}
\clearpage

\subsection{LoCoV1 and BEIR Document Length Distributions}

\begin{figure}[H]
   \centering
   \includegraphics[width=1.0\linewidth]{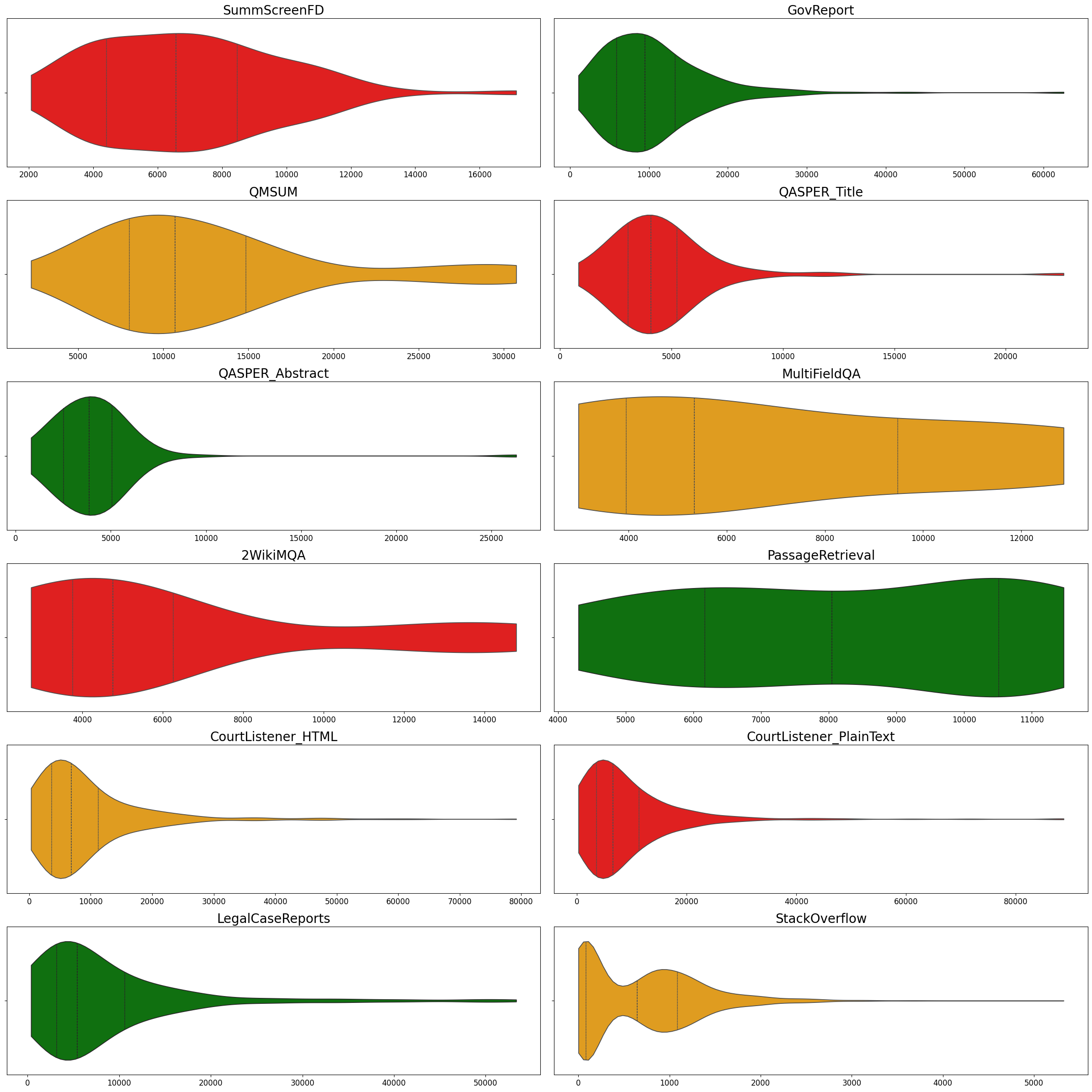}
   \caption{LoCoV1 Document Token Count Distributions.}
   \label{fig:loco_violins}
\end{figure}

\begin{figure}[H]
   \centering
   \includegraphics[width=1.0\linewidth]{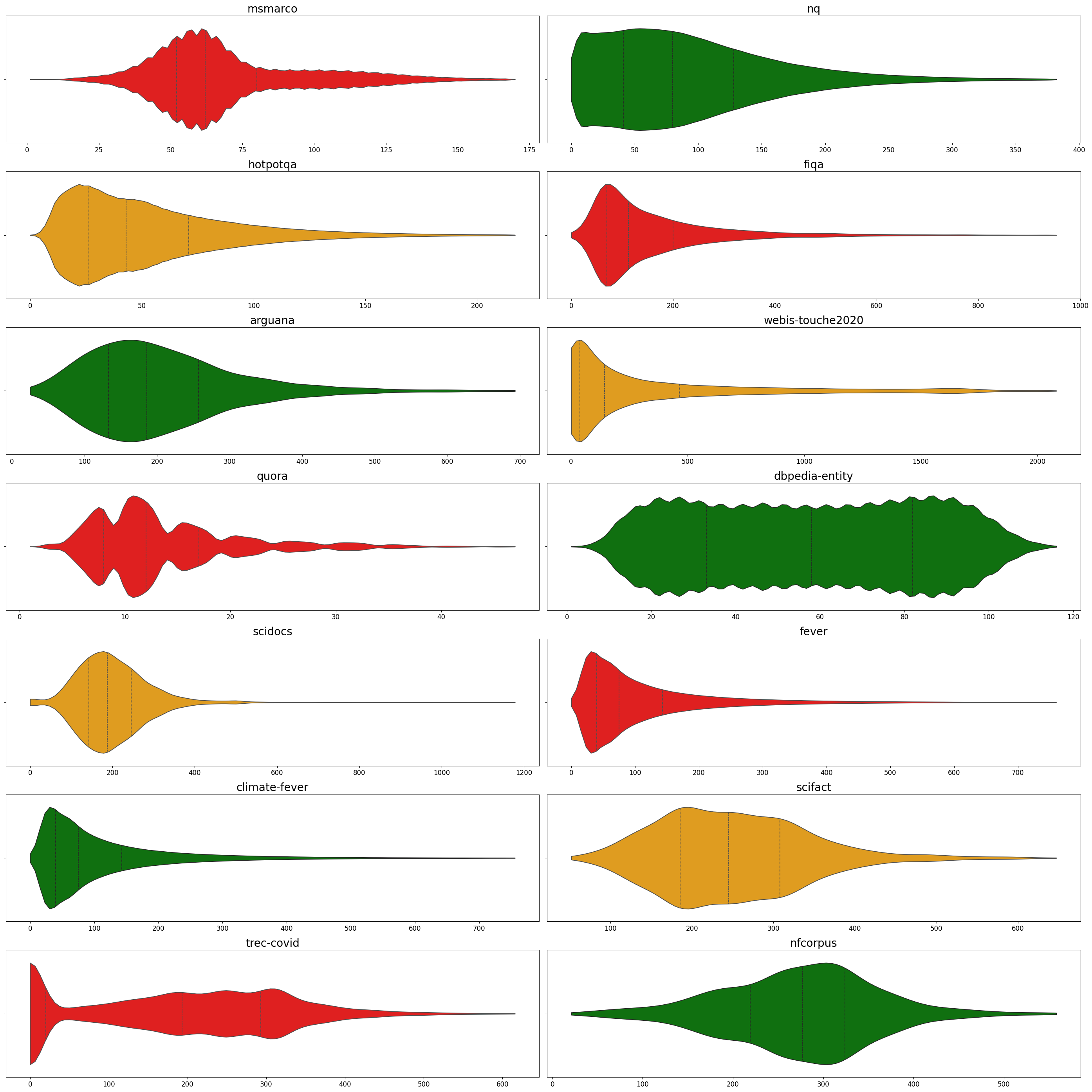}
   \caption{BEIR Document Token Count Distributions.}
   \label{fig:beir_violins}
\end{figure}
\clearpage

\subsection{LoCoV1 Performance Breakdown}

\begin{table}[H]
\centering
\begin{tabular}{ccccc}
\toprule
                                                              & \multicolumn{4}{c}{\textbf{\begin{tabular}[c]{@{}c@{}}nDCG@10 for \\ Document Subset\end{tabular}}}                                                                                      \\ \midrule
\textbf{Model}                                                & \textless{}2k & \begin{tabular}[c]{@{}c@{}}\textgreater{}2k\\ \textless{}8k\end{tabular} & \begin{tabular}[c]{@{}c@{}}\textgreater{}8k\\ \textless{}32k\end{tabular} & \textgreater{}32k \\ \midrule
\begin{tabular}[c]{@{}c@{}}BGE-Large\\ Zero-Shot\end{tabular} & 34.2          & 39.2                                                                     & 32.6                                                                      & 13.3              \\ \midrule
Mistral                                                       & 54.5          & 60.7                                                                     & 47.9                                                                      & 24.8              \\ \midrule
M2-BERT-128                                                   & 70.8          & 60.2                                                                     & 34.3                                                                      & 15.4              \\ \midrule
M2-BERT-2k                                                    & 63.9          & 68.1                                                                     & 47.9                                                                      & 24.1              \\ \midrule
M2-BERT-8k                                                    & 88.5          & 90.6                                                                     & 89.9                                                                      & 81.1              \\ \midrule
M2-BERT-32k                                                   & 90.4          & 93.1                                                                     & 94.4                                                                      & 86.1  \\ \bottomrule           
\end{tabular}
\caption{M2-BERT Encoder and Baseline Performances by Document Length. Queries are filtered by whether the token length of their answer documents are in the token range.}
\end{table}
\clearpage

\begin{table}[H]
\centering
\tiny
\setlength{\tabcolsep}{1.0pt}
\begin{tabular}{cll}
\toprule
\textbf{Model} & \multicolumn{1}{c}{\textbf{Query}}                                                                                                                                                                                                                                                                                                                                                                                                                                                                                                                                                                                                                                                                                                                                                                                                                                                                                        & \multicolumn{1}{c}{\textbf{Passage}}                                                                                                                                                                                                                                                                                                                                                                                                                                                                                                                                                                                                                                                                                                                                                                                                                                                                                                                                                                                                                                      \\ \midrule
\begin{tabular}[c]{@{}c@{}}BGE\\Large\end{tabular}      & \begin{tabular}[c]{@{}l@{}}This report discusses runaway and homeless \\youth, and the federal response to support this population. \\ There is no single definition of the terms "runaway youth"\\ or "homeless youth." However, both groups of youth \\ share the risk of not having adequate shelter and \\ other provisions, and may engage in harmful  \\ behaviors while away from a permanent home.\end{tabular}                                                                                                                                                                                                                                                                                                                                                                                                                                                                                                        & \begin{tabular}[c]{@{}l@{}}Running away from home is not a recent phenomenon. Folkloric heroes \\ Huckleberry Finn and Davy Crockett fled their abusive fathers to find \\ adventure and employment. Although some youth today also leave home \\ due to abuse and neglect, they often endure far more negative outcomes \\ than their romanticized counterparts from an earlier era. Without \\ adequate and safe shelter, runaway and homeless youth are vulnerable to \\ engaging in high-risk behaviors and further victimization. Youth who live \\ away from home for extended periods may become removed from...\end{tabular}                                                                                                                                                                                                                                                                                                                                                                                                                                       \\ \midrule
\begin{tabular}[c]{@{}c@{}}BGE\\Large\end{tabular}      & \begin{tabular}[c]{@{}l@{}}The professor thought it was possible to reduce the\\ effects of reverberation by removing the low-energy\\ segments. He thought a VAD-like approach would work. \\ This would make it so that the model was more \\likely to keep an echo than throw out speech.\end{tabular}                                                                                                                                                                                                                                                                                                                                                                                                                                                                                                                                                                                                                 & \textcolor{red}{\begin{tabular}[c]{@{}l@{}}Professor B: I think for two years we were two months , uh , away \\ from being done . \\ PhD A: And what was that , Morgan ? What project ?\\ Professor B: Uh , the , uh , TORRENT chip . \\ PhD A: Oh .\\ Professor B: Yeah . We were two \{disfmarker\} we were \{disfmarker\}\\ PhD C: Yeah .\\ Professor B: Uh , uh , we went through it...\end{tabular}}                                                                                                                                                                                                                                                                                                                                                                                                                                                                                                                  \\ \midrule
\begin{tabular}[c]{@{}c@{}}BGE\\Large\end{tabular}      & \begin{tabular}[c]{@{}l@{}}‘‘{[}i{]}n deciding cases . . . {[}j{]}urors are not expected to \\lay aside matters of common knowledge or their own \\observations and experiences, but rather, to apply\\ them to the facts as presented to arrive at an \\intelligent and correct conclusion’’ (internal \\quotation marks omitted)\end{tabular}                                                                                                                                                                                                                                                                                                                                                                                                                                                                                                                                                                         & \textcolor{red}{\begin{tabular}[c]{@{}l@{}}The “officially released” date that appears near the beginning of each \\ opinion is the date the opinion will be published in the Connecticut \\ Law Journal or the date it was released as a slip opinion. The operative \\ date for the beginning of all time periods for filing postopinion motions \\ and petitions for certification is the “officially released” date appearing \\ in the opinion. All opinions are subject to modification and technical \\ correction prior to official publication in the Connecticut Reports and \\ Connecticut Appellate Reports. In the event of discrepancies between...\end{tabular}}                                                                                                                                                                                                                                                                                                                                                                                                             \\ \midrule
\begin{tabular}[c]{@{}c@{}}E5\\Mistral\end{tabular}     & \begin{tabular}[c]{@{}l@{}}In this paper, we describe a new national language technology\\ programme for Icelandic. The programme,\\ which spans a period of five years, aims at making \\Icelandic usable in communication and interactions \\in the digital world, by developing accessible, open-\\source language resources and software. The research \\and development work within the programme is carried \\out by a consortium of universities, institutions, \\and private companies, with a strong emphasis on \\cooperation between academia and industries. Five \\core projects will be the main content of the \\programme: language resources, speech recognition, \\speech synthesis, machine translation, and spell and \\grammar checking. We also describe other national \\language technology programmes and give an overview \\over the history of language technology in \\Iceland.\end{tabular} & \begin{tabular}[c]{@{}l@{}}During the last decade, we have witnessed enormous advances in \\ language technology (LT). Applications that allow users to interact \\ with technology via spoken or written natural language are emerging \\ in all areas, and access to language resources and open-source software \\ libraries enables faster development for new domains and languages. \\ However, LT is highly language dependent and it takes considerable \\ resources to develop LT for new languages. The recent LT development \\ has focused on languages that have both a large number of speakers and \\ huge amounts of digitized language resources, like English, German, Spanish, \\ Japanese, etc. Other languages, that have few speakers and/or lack digitized \\ language resources, run the risk of being left behind. Icelandic is an example \\ of a language with almost a negligible number of speakers, in terms of...\end{tabular}                                                                                                              \\ \midrule
\begin{tabular}[c]{@{}c@{}}E5\\Mistral\end{tabular}     & Who was Brooksley Elizabeth's first husband?                                                                                                                                                                                                                                                                                                                                                                                                                                                                                                                                                                                                                                                                                                                                                                                                                                                                              & \begin{tabular}[c]{@{}l@{}}Brooksley Elizabeth Born (born August 27, 1940) is an American attorney \\ and former public official who, from August 26, 1996, to June 1, 1999, was \\ chair of the Commodity Futures Trading Commission (CFTC), the federal agency \\ which oversees the U.S. futures and commodity options markets. During her tenure \\ on the CFTC, Born lobbied Congress and the President to give the CFTC oversight \\ of off-exchange markets for derivatives, in addition to its role with respect to...\end{tabular}                                                                                                                                                                                                                                                                                                                                                        \\ \midrule
\begin{tabular}[c]{@{}c@{}}E5\\Mistral\end{tabular}     & \begin{tabular}[c]{@{}l@{}}Niles is scanning the society page when he sees a picture of \\Maris with another man. He plans to take an heiress \\on a date at a society event, the Snow Ball. He then \\realizes that he cannot dance but Daphne then offers \\to teach him. His date cancels, prompting Daphne to \\suggest that she go with him to the Ball. At the \\ball, Niles and Daphne dance, to show everyone there \\that he is not mourning his divorce. As they dance a \\tango, Niles declares that he adores Daphne, and...\end{tabular}                                                                                                                                                                                                                   & \textcolor{red}{\begin{tabular}[c]{@{}l@{}}
ACT ONE Scene One - KACL Frasier\textbackslash{}'s on\\ air at KACL and he\textbackslash{}'s running out\\ of time. But Roz still hands him over to his next caller.\textbackslash{}nFrasier: Well, \\we\textbackslash{}'ve got about thirty seconds. I \\think we\textbackslash{}'ve got time for one quick \\call. {[}presses button{]} Hello, Marlene, \\I\textbackslash{}'m listening.\textbackslash{}nMarlene: {[}v.o.{]} Oh my \\God, I\textbackslash{}'m really on?\textbackslash{}nFrasier: Yes, your problem,\\ please...\textbackslash{}nMarlene: {[}dog barking{]} \\Lucky, Lucky, get down. George, get the dog... \end{tabular}}                                      \\ \midrule
\begin{tabular}[c]{@{}c@{}}M2-BERT\\32k\end{tabular}    & {\begin{tabular}[c]{@{}l@{}} Which country Albertine, Baroness Staël\\ Von Holstein's father is from?\end{tabular}}                                                                                                                                                                                                                                                                                                                                                                                                                                                                                                                                                                                                                                                                                                                                                                                                                                                 & \begin{tabular}[c]{@{}l@{}}Passage 1:\textbackslash{}nAlbertine, baroness Staël von Holstein\textbackslash{}nHedvig Gustava Albertina, \\ Baroness de Staël-Holstein or simply Albertine (1797–1838), was the daughter \\ of Erik Magnus Staël von Holstein and Madame de Staël, the granddaughter of...\end{tabular}                                                                                                                                                \\ \midrule
\begin{tabular}[c]{@{}c@{}}M2-BERT\\32k\end{tabular}    & \begin{tabular}[c]{@{}l@{}}
The text is about Calvin Zabo, a biochemist who\\ becomes obsessed with the idea of transforming into\\ a superhuman form similar to the character Mr. Hyde\\ in Stevenson's novel. He robs his employers to fund\\ his experiments and seeks revenge on Donald Blake, a\\ doctor who refuses to give him a job. Zabo \\successfully creates a formula that transforms him\\ into a Hulk-like creature called Mister Hyde. Hyde\\ attempts to kill Blake, but Blake transforms into Thor\\ and survives. \end{tabular}                                                                                                                                                                                          & \begin{tabular}[c]{@{}l@{}}Paragraph 1: With very few feature films made in Canada at all prior \\to the 1960s, in some years no Film of the Year winner was named at all, \\with the awards for Best  Short Film or Best Amateur Film instead \\constituting the highest honour given to a  film that year.\\ Even the award for Film of the Year, when presented at all, \\ often also went to a short film. The awards were\\ also almost totally dominated by the National Film Board,\\ to the point that independent filmmakers sometimes alleged a systemic \\ bias which was itself a contributing factor to the difficulty\\ of building a sustainable \end{tabular}                                                                                                                                                                                                               \\ \midrule
\begin{tabular}[c]{@{}c@{}}M2-BERT\\32k\end{tabular}    & \begin{tabular}[c]{@{}l@{}}“{[}T{]}he rules of criminal procedure\\ require the appointment of counsel in PCRA proceedings.”\end{tabular}                                                                                                                                                                                                                                                                                                                                                                                                                                                                                                                                                                                                                                                                                                                                                                                & \begin{tabular}[c]{@{}l@{}}J-S79022-17\textbackslash{}n\textbackslash{}n\textbackslash{}nNON-PRECEDENTIAL DECISION -\\ SEE SUPERIOR COURT I.O.P. 65.37\textbackslash{}n\textbackslash{}nCOMMONWEALTH OF\\ PENNSYLVANIA : IN THE SUPERIOR COURT OF\textbackslash{}n  : PENNSYLVANIA\textbackslash{}n\\ :\textbackslash{}n v. VERNELL MORRIS            Appellant : No. 3731 EDA 2016\textbackslash{}n\textbackslash{}n              \\ Appeal from the PCRA Order November 3, 2016\textbackslash{}n\\ In the Court of Common Pleas of Philadelphia County Criminal Division at\textbackslash{}n                  \\ No(s): CP-51-CR-1113151-1992\textbackslash{}n\textbackslash{}n\textbackslash{}nBEFORE: GANTMAN, P.J.,...\end{tabular} \\ \bottomrule
\end{tabular}
\caption{\textbf{LoCoV1 Performance Analysis by Model}: Passages that aren't highlighted were retrieved successfully while passages highlighted in red were not successfully retrieved.
Retrieval success is defined as whether it was retrieved in the first 10 passages.}
\end{table}

\appendix

\end{document}